\documentclass[%
 reprint,
superscriptaddress,
 amsmath,amssymb,
 aps,
]{revtex4-2}

\usepackage{graphicx}
\usepackage{dcolumn}
\usepackage{bm}
\usepackage{color}
\usepackage{amsmath}
\usepackage{graphics}
\usepackage{gensymb}
\usepackage{subfigure}
\usepackage{pdftexcmds}
\usepackage{ifpdf}
\usepackage{amssymb}
\usepackage{dsfont}
\usepackage{soul}

\def \be {\begin{equation}}
\def \ee {\end{equation}}
\def \bea {\begin{eqnarray}}
\def \eea {\end{eqnarray}}

\begin{document}

\preprint{APS/123-QED}

\title{Chiral helimagnetism and stability of magnetic textures in MnNb$_3$S$_6$}

\author{S. A. Osorio}
\affiliation{ 
Instituto de Nanociencia y Nanotecnolog\'ia (CNEA-CONICET), Nodo Bariloche, Av. Bustillo 9500 (R8402AGP), S. C. de Bariloche, R\'io Negro, Argentina
}
\affiliation{
Gerencia de F\'isica,  Centro At\'omico Bariloche, Av. Bustillo 9500 (R8402AGP), S. C. de Bariloche, R\'io Negro, Argentina
}

\author{V. Laliena}
\affiliation{
Department of Applied Mathematics, University of Zaragoza, C/ Mar\'ia de Luna 3, 50018 Zaragoza, Spain
}

\author{J. Campo}
\affiliation{
Aragon Nanoscience and Materials Institute (CSIC-University of Zaragoza) and Condensed Matter Physics Department, University of Zaragoza, C/ Pedro Cerbuna 12, 50009 Zaragoza, Spain
}

\author{S. Bustingorry}
\affiliation{ 
Instituto de Nanociencia y Nanotecnolog\'ia (CNEA-CONICET), Nodo Bariloche, Av. Bustillo 9500 (R8402AGP), S. C. de Bariloche, R\'io Negro, Argentina
}
\affiliation{
Gerencia de F\'isica,  Centro At\'omico Bariloche, Av. Bustillo 9500 (R8402AGP), S. C. de Bariloche, R\'io Negro, Argentina
}
\affiliation{
Aragon Nanoscience and Materials Institute (CSIC-University of Zaragoza) and Condensed Matter Physics Department, University of Zaragoza, C/ Pedro Cerbuna 12, 50009 Zaragoza, Spain
}

\date{\today}

\begin{abstract}
We analyze the nature of the modulated magnetic states in a micromagnetic model for the monoaxial chiral magnet MnNb$_3$S$_6$ for which the Dzyaloshinskii-Moriya interaction and the dipolar interaction compete evenly.
We show that the interplay between these interactions lead to a complex phase diagram including ferromagnetic states, fan-like states, in-plane stripes patterns and chiral soliton lattices. In particular, stripe patterns and chiral soliton lattices comprise non-trivial topological states with fixed chirality.
The obtained phase diagram exhibits strong dependency on the thickness and on the strength of the Dzyaloshinskii-Moriya interaction. Our results can help to understand the magnetic properties of systems such as MnNb$_3$S$_6$.
\end{abstract}

\maketitle


%
\section{Introduction}
\label{sec:intro}
In bulk monoaxial chiral helimagnets the physics of the system is usually dominated by Dzyaloshinskii-Moriya interactions (DMI)~\cite{Dzyal58,moriya1960new}, as is the case of the compound CrNb$_3$S$_6$~\cite{Dzyal64,Moriya82,Miyadai83,izyumov1984modulated,kishine2005synthesis,togawa2012chiral,Kousaka16}. The magnetic properties of the monoaxial chiral helimagnets are dominated by the presence of a chiral soliton lattice (CSL) when a magnetic field is applied perpendicular to the chiral axis~\cite{Dzyal64, Miyadai83, izyumov1984modulated, Zheludev97,kishine2005synthesis,togawa2012chiral,Ghimire13,Togawa13,Chapman14,Kishine15,Togawa16,Shinozaki16,Nishikawa16,Laliena16a,Laliena16b,Tsuruta16,han2017tricritical,yonemura2017magnetic,clements2017critical,Laliena17a,Masaki18,Laliena18a,aoki2019anomalous,honda2020topological,Masaki20}. In this case the dipolar interaction, being much weaker than the DMI, does not play an essential role for many practical situations and therefore it is not taken into account in most of the theoretical studies about these compounds.
However, strong dipolar effects can lead to a plethora of modulated phases and blur the previous picture in rather subtle ways.
Recent experimental results suggest that this scenario takes place in the compound MnNb$_3$S$_6$ \cite{Karna2021}. 
Theoretical and experimental works on MnNb$_3$S$_6$ suggest the presence of a DMI~\cite{mankovsky2016electronic} and magnetic properties analogous to those observed in the isostructural compound CrNb$_3$S$_6$~\cite{kousaka2009chiral,karna2019consequences,dai2019critical,ohkuma2022nonequilibrium,li2023temperature}. However, the presence of helimagnetic order in MnNb$_3$S$_6$ and the role of the dipolar interaction are still a matter of debate and requires further studies~\cite{karna2019consequences,hall2022comparative}. 

The purpose of this work is to analyze theoretically the nature of the magnetic states within the scenario presented in Ref. \onlinecite{Karna2021} in which both interactions, the DMI and the dipolar interaction, compete evenly.
We use extensive micromagnetic simulations to uncover the effect of the interplay between the DMI and the dipolar interaction. We explore how the ground state of the magnetic system can change with the thickness of the sample and the external magnetic field perpendicular to it.
We show that the phase diagram can exhibit a strong dependence on the thickness of samples which may hinder the unambiguous identification of the underlying magnetic textures.
We also analyze the magnetic configurations to trace back the distinctive hallmark of each magnetic texture. For these reasons our work could be helpful in the interpretation of experimental studies of the magnetic properties of MnNb$_3$S$_6$ and similar compounds in which the DMI and the dipolar interaction are both relevant.

We start the article by introducing the model and the methods. Then we discuss the energy landscape of modulated states as a function of their period, and the main effects of the thickness of the sample and the magnetic field perpendicular to the film. Then we study the equilibrium phase diagram at zero temperature as a function of the thickness of the system and the external magnetic field, and analyze the main properties of each magnetic configuration. We continue with an analysis of the metastability limits of the modulated phases. Finally we study how the intensity of the DMI affects the main properties of the modulated states.
We end our work with a summary and a discussion about the experimental implications of the obtained results.


%
\section{Methods}
\label{sec:methods}

To describe the monoaxial helimagnet we consider the following effective model for the energy density of the system $e(\bm{r})$:
\bea
\nonumber
e(\bm{r})&=&A\sum_{i}\left(\partial_{i}\bm{n}\right)^2-D\bm{\hat{z}}\cdot\left(\bm{n}\times\partial_{z}\bm{n}\right)-K n_{z}^{2}\\
&&-M_{\mathrm{S}}\bm{B}\cdot\bm{n}-\frac{1}{2}\mu_{0}M_{\mathrm{S}}\bm{H}_{\mathrm{d}}\cdot\bm{n},
\label{eq:model}
\eea
where $\bm{n}(\bm{r}) = \bm{M}(\bm{r})/M_\mathrm{S}$ stands for the local normalized magnetization, with $M_\mathrm{S}$ the saturation magnetization.
The first term in Eq.~\eqref{eq:model} represents the exchange interaction with stiffness constant $A$, the second term is the DMI with intensity $D$, the third term is the easy-plane anisotropy with strength $K<0$, and $\bm{B}$ is the external magnetic field. The effects of the dipolar interaction are included in the last term, with the dipolar field $\bm{H}_{\mathrm{d}}$. The index $i$ runs over $x,y,z$ and the chiral axis is along $\bm{\hat{z}}$. The total energy of the system is given by $E[\bm{n}(\bm{r})]=\int d^{3}\bm{r} e(\bm{r})$.

It was recently shown that the MnNb$_3$S$_6$ compound presents a strong magnetic easy-plane  anisotropy (perpendicular to the chiral axis of the compound), and its critical properties are compatible with a three-dimensional Heisenberg model \cite{dai2019critical}. Thus the energy density introduced in Eq. \eqref{eq:model} can be used to model the compound MnNb$_3$S$_6$. Following the values reported by Karna et al \cite{Karna2021} we take $A=1.0$ pJ$/$m, $D=50$ $\mu$J$/$m$^{2}$, $K=-322$ kJ$/$m$^3$ and $M_{\mathrm{S}}=215$ kA$/$m. It is important to compare these parameters with those for CrNb$_3$S$_6$ \cite{victor2020dynamics, Osorio2021}. On one side the value of $D$ for MnNb$_3$S$_6$ is smaller than that for CrNb$_3$S$_6$ by a factor $\sim 1/7$. On the other side, the saturation magnetization of MnNb$_3$S$_6$ is larger than $M_{\mathrm{S}}$ for CrNb$_3$S$_6$ by a factor $\sim 5/3$, which suggest that the dipolar term, proportional to $M_\mathrm{S}^2$, plays a crucial role. Then $\mu_{0}M_\mathrm{S}^{2}/D$, which measures the relative strength of dipolar over DMI terms, is 20 times larger for MnNb$_3$S$_6$.

We use the MuMax software to perform micromagnetic simulations \cite{MuMax3,Leliaert18}. 
We consider an infinite system in the $x$ and $z$ directions, and finite in the $y$ direction, with thickness $t$. The system is discretized using cubic cells with edges of length $\Delta = 4$ nm. The infinite dimensions are simulated, as usual, by imposing periodic boundary conditions in the $x$ and $z$ directions. Due to the monoaxial nature of the DMI we do not expect any modulation in the $x$ direction (for the infinite system) and thus we fix the system size in the $x$ direction to $w=40$ nm. However, we expect modulations in the $z$ direction, and therefore the system size in the $z$ direction, $L$, is a variational parameter which has to be fixed by energy minimization (see below). In practice we let $L$ vary in the range $40$ nm $< L < 1000$ nm. The goal is to study the dipolar effects and analyze the equilibrium and metastable states of the system as a function of the thickness, $t$, and of the applied field, $\bm{B} = B \bm{\hat{y}}$, perpendicular to the system plane. To this end we let $t$ vary in the range $12$ nm $\leq t \leq 360$ nm and $B$ in the range $0$ mT $\leq B\leq 240$ mT.


In order to find the magnetic configurations with minimum energy, we initialize the system in four different configurations, minimizing the possibility to be trapped in a metastable configuration. Two uniform initial configurations were considered: the in-plane uniform configuration along $+x$ that we call UX, and the out-of-plane uniform configuration along $+y$ that we call UY. Besides, two modulated phases (along the chiral axis) corresponding to two different topological sectors were used as initial configurations: along the $z$ direction the system was split into two in-plane domains with the magnetization pointing in the $+x$ and $-x$ opposite directions, separated by two Bloch domain walls.
In the heterochiral initial state (HEC) the two Bloch domain walls have the magnetization at the center of the domain walls pointing in the same direction ($+y$), thus having opposite chirality. In the homochiral initial state (HOC), the domain walls have the same chirality, i.e. with the magnetization at the center pointing in opposite $\pm y$ directions. Since the DMI favors one chiral state over the other, in the HEC state each domain wall has different energy, while in the HOC the chirality degeneracy is broken by selecting the chirality favored by the DMI term. These different initial conditions serves to arrive at different final states when the system is let to relax. The restriction to the previous selected states in the exploration of the phase diagram is motivated by the experimental results recently found in the compound MnNb$_3$S$_{6}$~\cite{Karna2021}.

For the HEC and HOC states, the distance between the center of consecutive domain walls is $L/2$. For fixed values of $t$, $B$, and $L$ the system relaxes to a minimal energy configuration. 
Since the energy density depends on $L$, we determine the equilibrium state by letting $L$ vary until the energy density attains a minimum value.
The results of this relaxation protocol permits us to identify low energy configuration states, as described in the next section.

\section{Results}
\label{sec:results}
\subsection{Energy curves}
\label{sec:landscapes}

To analyze the results we plot the energy density $\mathcal{E}(L)=E(L)/(t\times w\times L)$ as a function of $L$. We call these plots the energy curves and are obtained after a relaxation of the system from each of the four initial conditions.

The state obtained by relaxation depends, in general, on the starting point: UX, UY, HOC and HEC.
The energy density obtained after the relaxation from the uniform initial states, UX and UY, depends strongly on $L$ for $L\lesssim 100$ nm. This is due to the finite size effects introduced in computing the dipolar contribution in small sized samples. On the opposite extreme, large values of $L$, the energy density presents an asymptotic behavior to a constant value reflecting the translational invariance of a true uniform state. We assign this asymptotic value to the energy of the uniform states.

\begin{figure}[htb]
\subfigure{
\includegraphics[width=4cm]{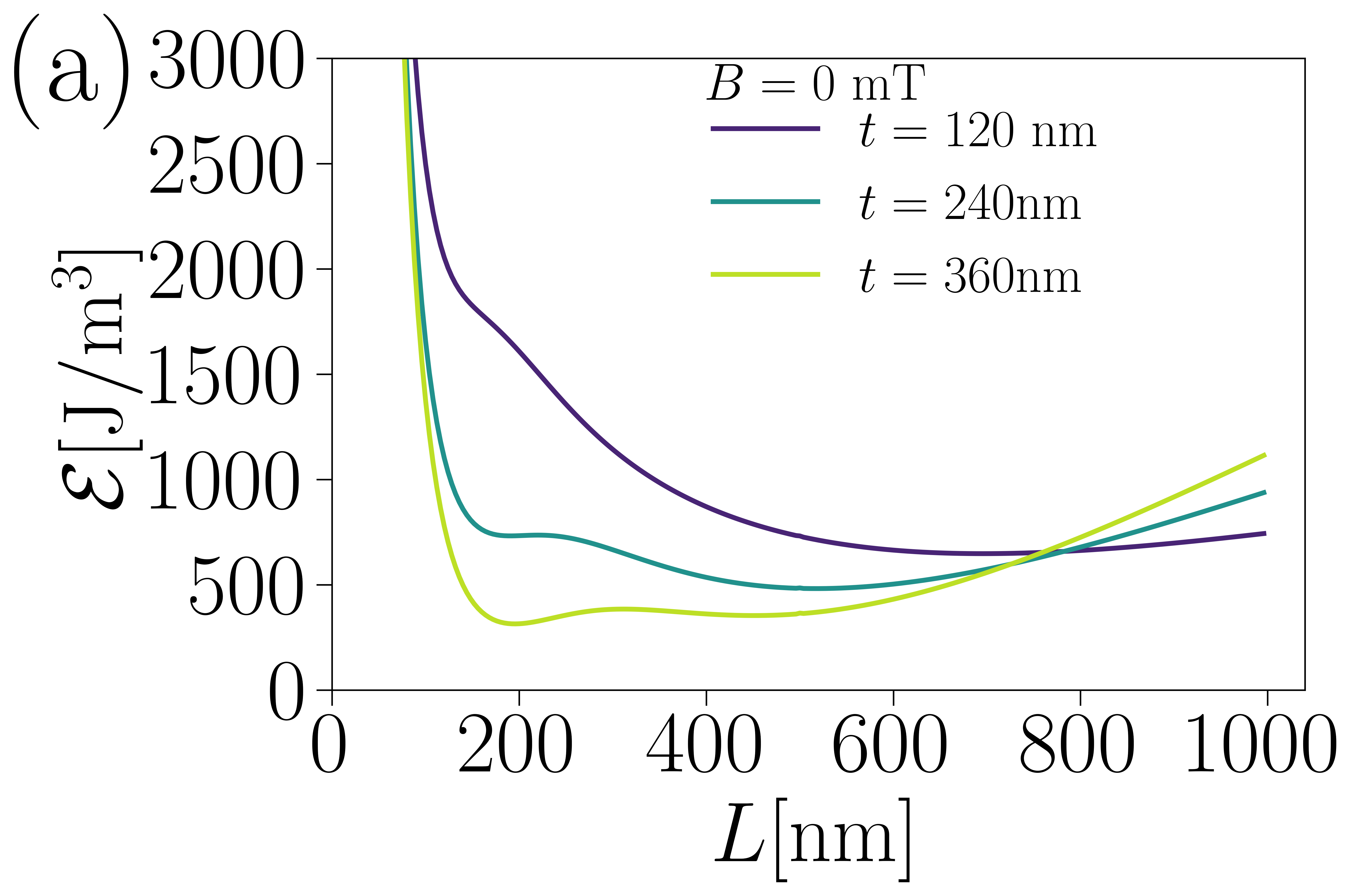}
}
\subfigure{
\includegraphics[width=4cm]{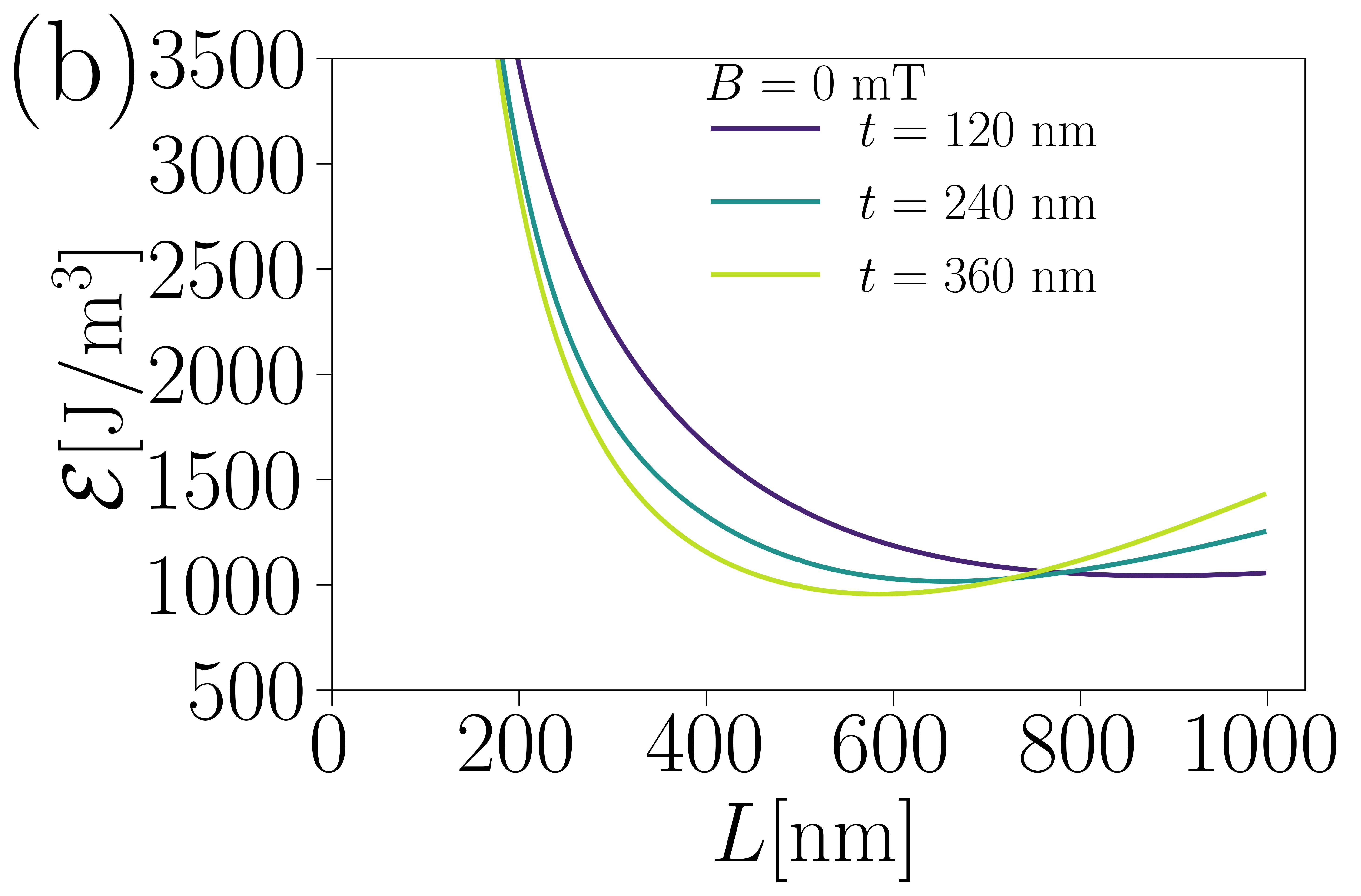}
}
\subfigure{
\includegraphics[width=4cm]{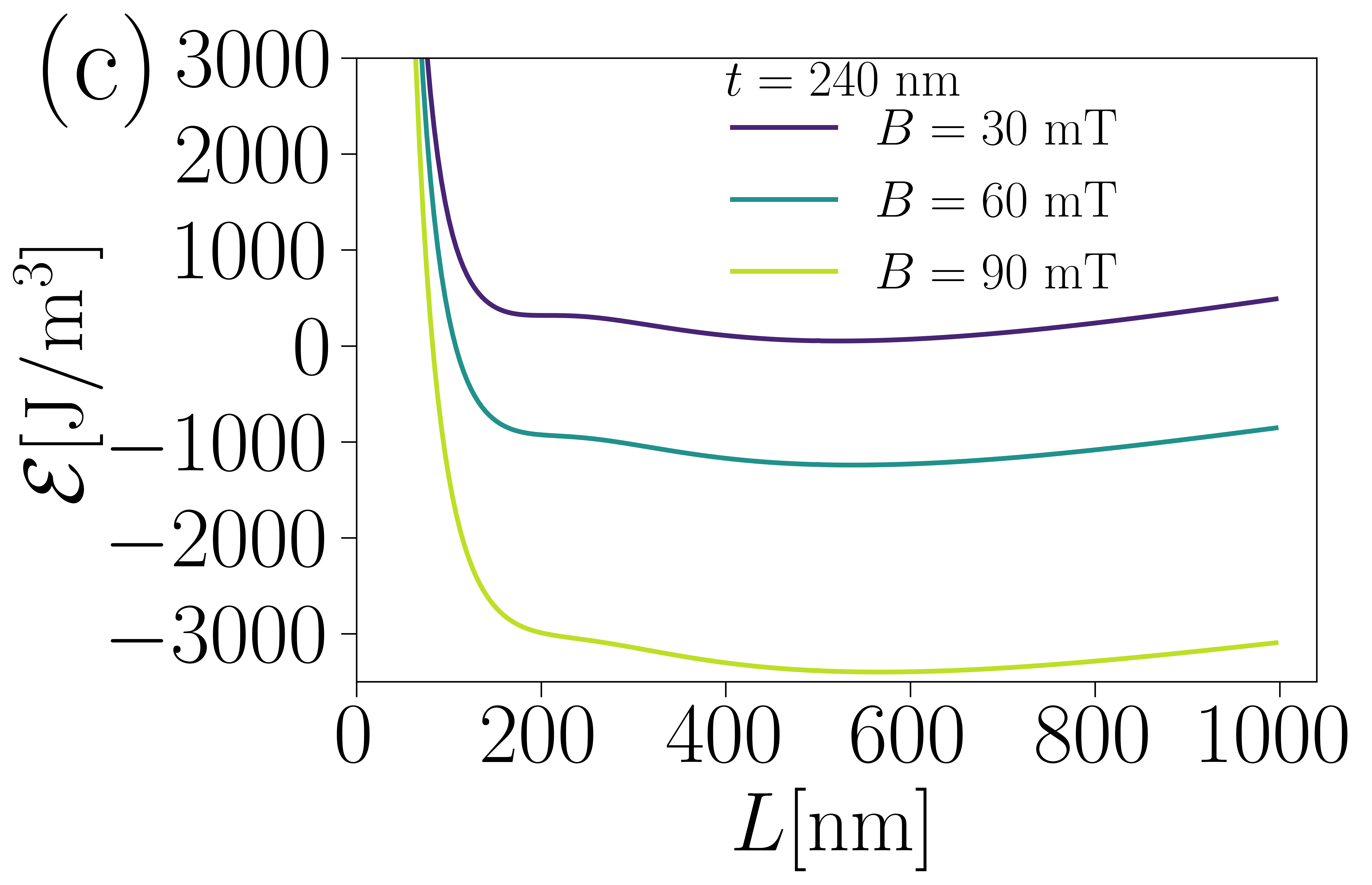}
}
\subfigure{
\includegraphics[width=4cm]{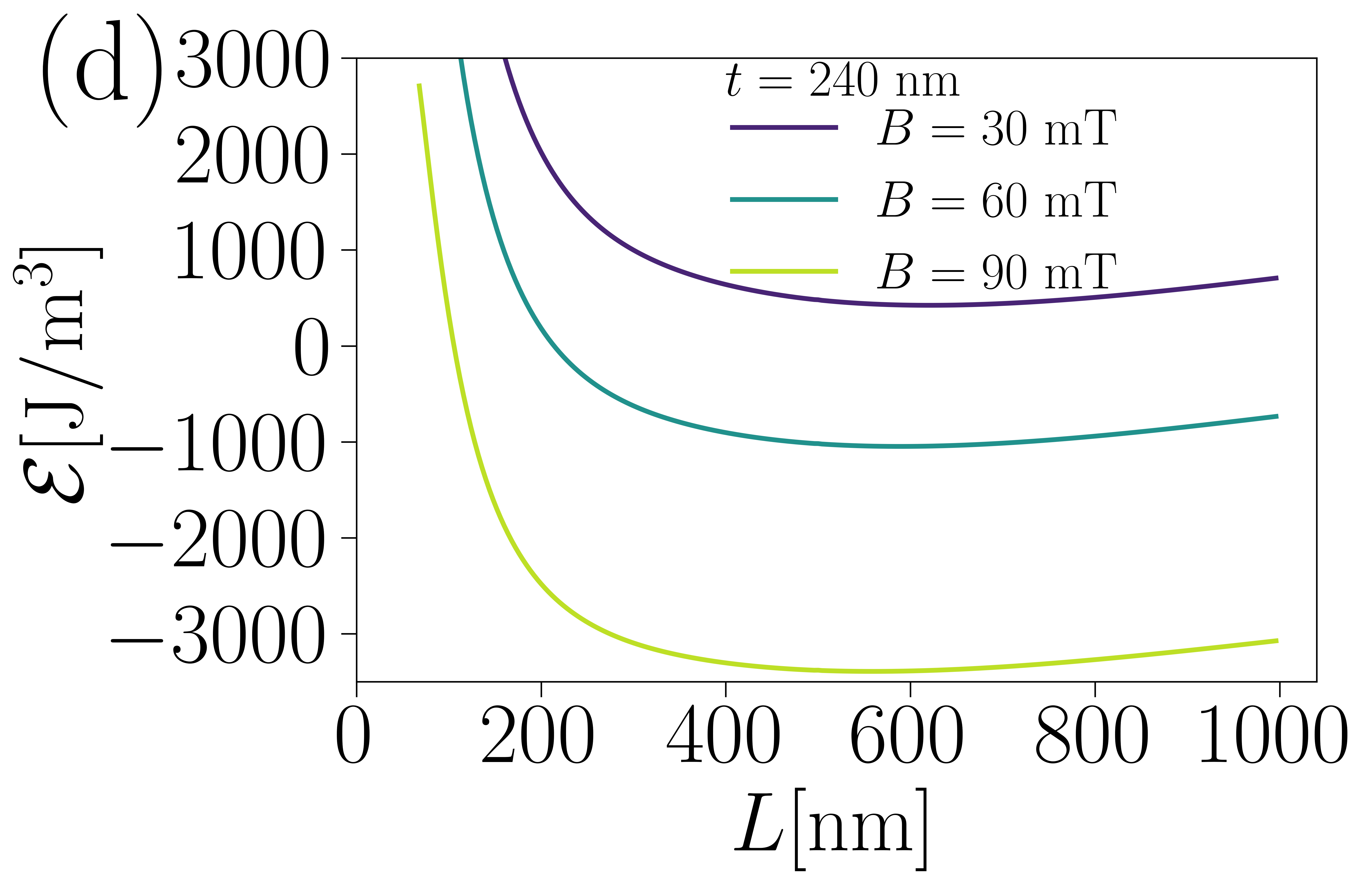}
}
\caption{
Top: Energy density as a function of the the period of the magnetic texture for $B=0$ mT and three values of the sample thickness $t=120$, $240$ and $360$ nm (in blue, watery green and green respectively), for the state that results after a relaxation from the initial condition corresponding to the HOC configuration in a) and to the HEC in b).
Bottom: Energy density as a function of the the period of the magnetic texture for $t=240$ nm and three values of the magnetic field $B=30$, $60$ and $90$ mT (in blue, watery green and green respectively), for the state that results after a relaxation from the initial condition corresponding to the HOC configuration in (c) and to the HEC in (d).
}
\label{fig:land}
\end{figure}

\begin{figure}[htb]
\includegraphics[width=9cm]{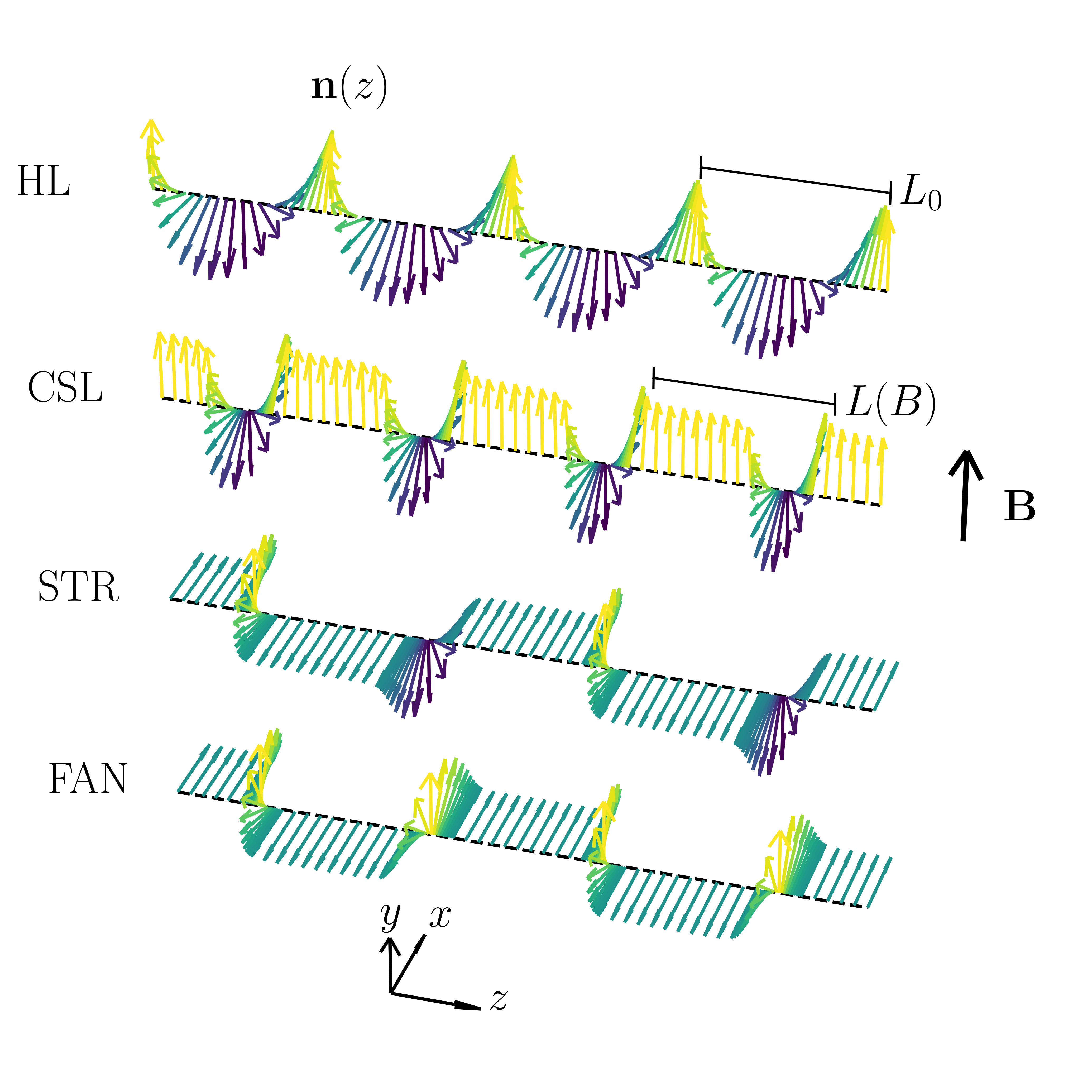}
\caption{Magnetization configurations: The helical (HL) state in a monoaxial chiral helimagnet at $B = 0$, the chiral soliton lattice (CSL) for $B \neq 0$, the stripes (STR) and the fan-like state (FAN). The color represents the $n_{y}$ component of the magnetization field: in yellow (blue) $n_{y}= +1$ ($-1$). The chiral axis is parallel to $\bm{\hat{z}}$.
}
\label{fig:schemes}
\end{figure}

The curves of energy densities versus $L$ obtained by relaxation from modulated states (HOC or HEC) show minima whose location and depth depend on the externally controlled parameters $t$ and $B$ (Fig. \ref{fig:land}). 
The energy curve resulting from the HOC initial condition (Fig. \ref{fig:land} (a), (c))exhibits two local minima located at different length scales, the first at $L\approx200$ nm and the second at $L\approx500$ nm. The magnetization fields present a different structure on each minimum. Around the first minimum the magnetization field of a slab well inside the bulk resembles that of a conventional helical (HL) state for $B=0$, or a CSL for $B\neq0$ (Fig. \ref{fig:schemes}). Therefore, we generically refer to the state associated with the first local minimum as a CSL state. For the second minimum the magnetization field inside the bulk has a strong in-plane component and resembles the structure of a magnetic stripe (STR) with in-plane domains. Thus, we associate the second minima to a STR state. Finally, the local minima in the energy obtained from the HEC initial configuration (Fig. \ref{fig:land} (b), (d)) corresponds to the fan-like state (FAN) shown in Fig. \ref{fig:schemes}.
It is important to note that the STR and CSL can be classified as topologically non trivial because the magnetization field winds an integer number of times around the chiral axis. The FAN state, however, is topologically trivial since the magnetization field oscillates back and forth without completing a whole turning around the chiral axis.

%
%

The energy curves change with $t$ and $B$, as shown in Fig. \ref{fig:land} for the HEC and the HOC states.
Thus, we study, in the next section, the equilibrium phase diagram for different values of the magnetic field $B$ and the sample thickness $t$. To this end, at each value of $B$ and $t$ we sweep the energy density for the values of $L$ within the range $40$ nm $\leq L\leq1000$ nm (with increments $\Delta L = 4$ nm). 

%
%

%
\subsection{The equilibrium phase diagram}
\label{sec:equilibrium}

To obtain the phase diagram we sweep the $(B,t)$ plane in the range $0$ mT $\leq B\leq 240$ mT and $12$ nm $\leq t \leq 360$ nm with increments $\Delta B = 6$ mT and $\Delta t = 12$ nm, respectively. For each $t$ and $B$, the equilibrium state corresponds to the absolute minimum of the whole set of energy curves. Therefore, at each point we repeat the protocol described above, which serves to determine the equilibrium state limited to our Ansätze for the initial states.


%
\begin{figure}[htb]
\includegraphics[width=7cm]{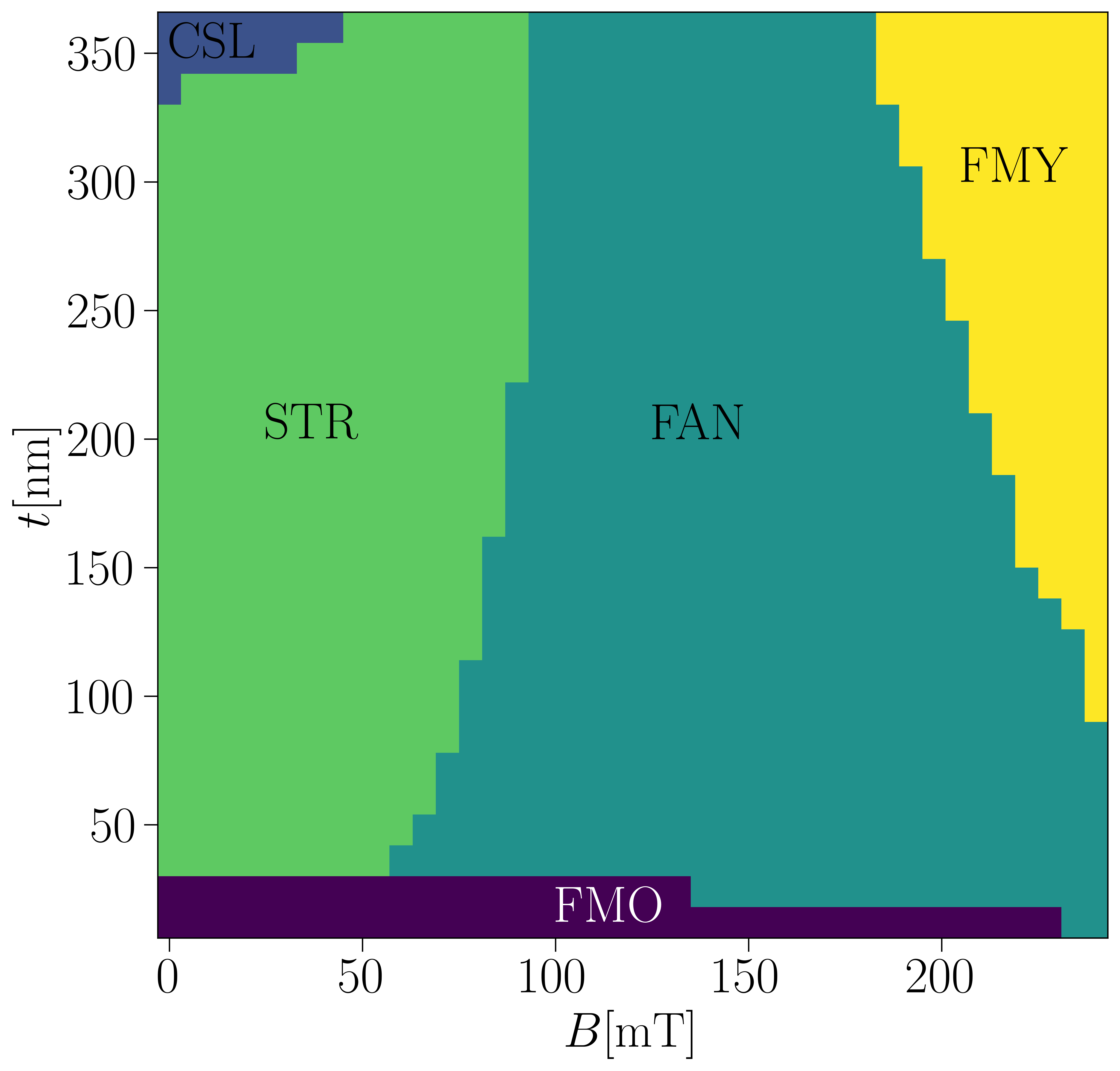}
\caption{
Equilibrium phase diagram in the $t-B$ space. There are two uniform states, the FMO and the FMY states, and three modulated states, the CSL, STR and FAN states.
}
\label{fig:diag_eq}
\end{figure}

\begin{figure*}[ht]
\includegraphics[width=18cm]{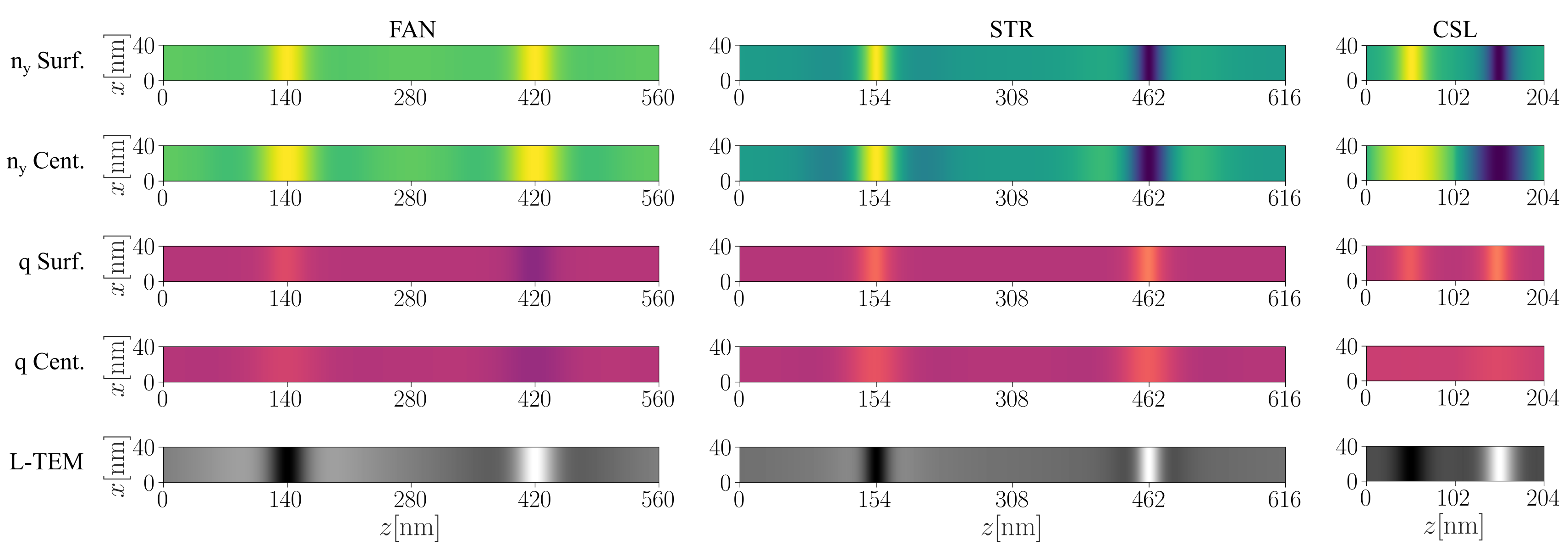}
\caption{Configuration of the magnetic field across one period of the three modulated states, the FAN ($t=204$ nm, $B=120$ mT), the STR ($t=156$ nm, $B=24$ mT) and the CSL ($t=360$ nm, $B=36$ mT). 
From top to bottom, the first and second rows represent the $n_{y}$ component of the magnetization field on the surface and at the center of the system, respectively. The color gradient goes from yellow ($n_{y}= +1$) to blue ($n_{y}= -1$).
The third and fourth rows represent the winding density $q$ of the magnetization field on the surface and at the center of the system, respectively. From purple to pink $-1 \leq q < 0$, and from pink to yellow $0 < q \leq 1$.
The fifth row represents the L-TEM intensities reconstructed  from the magnetization field.
}
\label{fig:config}
\end{figure*}

The phase diagram is shown in Fig.~\ref{fig:diag_eq}. We distinguish five regions in it.
For thin samples $t\lesssim 24$ nm the state corresponds to an oblique uniform state (FMO). The direction of the magnetization lies along the $x$ axis at zero field. As the magnetic field grows, the magnetization field gradually departs from the $xz$ plane to develop a transversal component along the magnetic field. Clearly, this is due to the competition between the Zeeman and dipolar interactions. For thicker systems, $t\gtrsim 100$ nm, and at magnetic field high enough, $B\gtrsim 175$ mT, the system is found in the uniform field polarized state along the magnetic field (FMY). This corresponds to the region in the upper right corner of the equilibrium phase diagram.

The CSL state is the equilibrium state for sufficiently large $t$ and low enough $B$. Only a small portion of this region appears in Fig.~\ref{fig:diag_eq}. This corresponds to weak fields $B\lesssim 50$ mT and thick samples $t\gtrsim 330$ nm. The CSL phase is surrounded by a region in which the STR state is the equilibrium state. This phase covers most of the sample thicknesses in a region ranging from zero to intermediate magnetic fields $B\lesssim 90$ mT. Finally, the FAN state is the equilibrium state for higher values of the magnetic field $B\gtrsim 90$ mT and thicknesses ranging from a few tens nanometers to more than 360 nm. We stress that there is a critical thickness $t_{\mathrm{CSL}}\approx 348$ nm for the CSL state to be stable. Below $t_{\mathrm{CSL}}$, the CSL cannot be an equilibrium state since the dipolar term favors in-plane magnetization directions for thinner samples. It is also important to emphasize that the FAN state cannot exist as a stable state for magnetic fields below $B_{\mathrm{FAN}}\approx 90$ mT. Above $B_{\mathrm{FAN}}$ the Zeeman energy overcomes the DMI for domain walls with a component $n_{y} < 0$, thus reversing the domain wall chirality and leading to the transition from STR to FAN.

To elucidate the structure of the magnetic state in each phase in Fig. \ref{fig:diag_eq} we analyze the magnetization field of the system. In particular we analyze how this structure changes from the center to the surface of the system. To this end we consider the $n_{y}$ component of the magnetization field and the winding density $q$ which is a scalar quantity defined as:
\be
q(\bm{r})=\left[\bm{n}_{\perp}(\bm{r})\times\bm{n}_{\perp}(\bm{r}+\Delta\bm{\hat{z}})\right]\cdot\bm{\hat{z}}.
\label{eq:chiral}
\ee
This field provides information about the rotation around the chiral axis $\bm{\hat{z}}$ of the transverse component of the magnetization field ($\bm{n}_{\perp}(\bm{r})\cdot\bm{\hat{z}}=0$ and $\bm{n}_{\perp}(\bm{r})\cdot\bm{n}_{\perp}(\bm{r})=1$) when we move along $\bm{\hat{z}}$. The quantity $q(\bm{r})$ can be interpreted as a measure of the local chirality.

Figure~\ref{fig:config} shows snapshots of the three modulated states encountered in the simulations. The first and second rows correspond to the $n_y$ component of the magnetization field on the surface and at the center of the system, respectively, for each state. The yellow (blue) regions correspond to $n_{y}=+1$ ($n_{y}=-1$), and these regions are separated by domains in which the magnetization lays within the $xz$ plane, represented in green color. The configurations at the center of the system resembles those shown in Fig. ~\ref{fig:schemes} which we use to identify the equilibrium states with FAN, STR and CSL states.
We observe that the configuration of the FAN and the STR states on the surface and at the center of the system exhibits a slight change, but they retain their main characteristics. On the contrary, the structure of the CSL state presents a strong variation along the direction perpendicular to the system surfaces. As can be seen in Fig.~\ref{fig:config}, the CSL state actually resembles a CSL in the center and a STR  (with the periodicity of the CSL) on the surface. 
The third and fourth rows represent the winding density $q$ (which provides the local chirality of the magnetization field), on the surface and at the center of the system, respectively. Two important things should be noticed from this figures. First, the FAN state is made of heterochiral structures signaled by the yellowish and purple fringes with opposite chiralities. The STR and the CSL states, however, are comprised of homochiral entities represented by yellow regions (the absence of purple regions means that $q > 0$). Second, the degree of localization of the chirality seems to be stronger in the FAN and the STR states than in the CSL state, as in this state the winding density is almost uniform.
Let us stress that the period of the FAN and the STR states are roughly similar, $L\approx550$ nm for the FAN state and $L\approx600$ nm for the STR state. These lengths are well above the period of the CSL state, for which $L\approx200$ nm.

Finally, in the fifth row of Fig.~\ref{fig:config} we show the expected L-TEM images reconstructed from the magnetization states. The L-TEM figures were obtained using the Ubermag package~\cite{beg2022} for an accelerating voltage of $\Delta V=300$ kV and a defocus length $\Delta f = 800$ nm, which are values similar to those used in the experiments~\cite{togawa2012chiral, togawa2015magnetic, Karna2021}. We can observe that there are light and dark fringes that exhibit a similar structure across the three different states. For the FAN and STR states, the light and dark fringes are well separated by a gray background. The size of the background is many times larger than the width of the light and dark fringes. In the CSL however, the size of the background is similar to the width of the light and dark fringes. The simulated L-TEM for the STR and the FAN states are roughly consistent with those obtained experimentally \cite{Karna2021,hall2022comparative} for MnNb$_3$S$_6$.

\begin{figure}[htb!]
\subfigure{
\includegraphics[width=8cm]{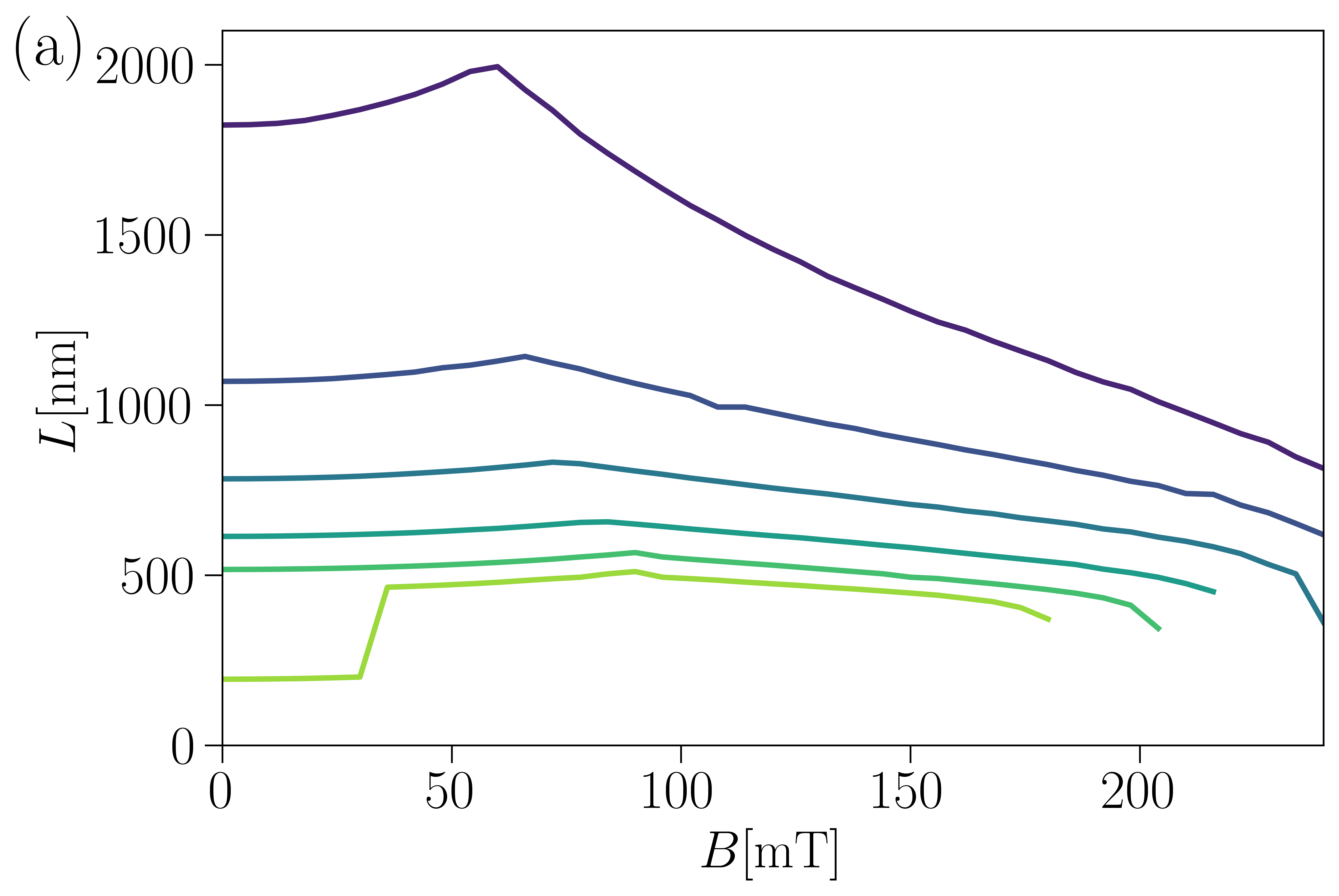}
}
\subfigure{
\includegraphics[width=8cm]{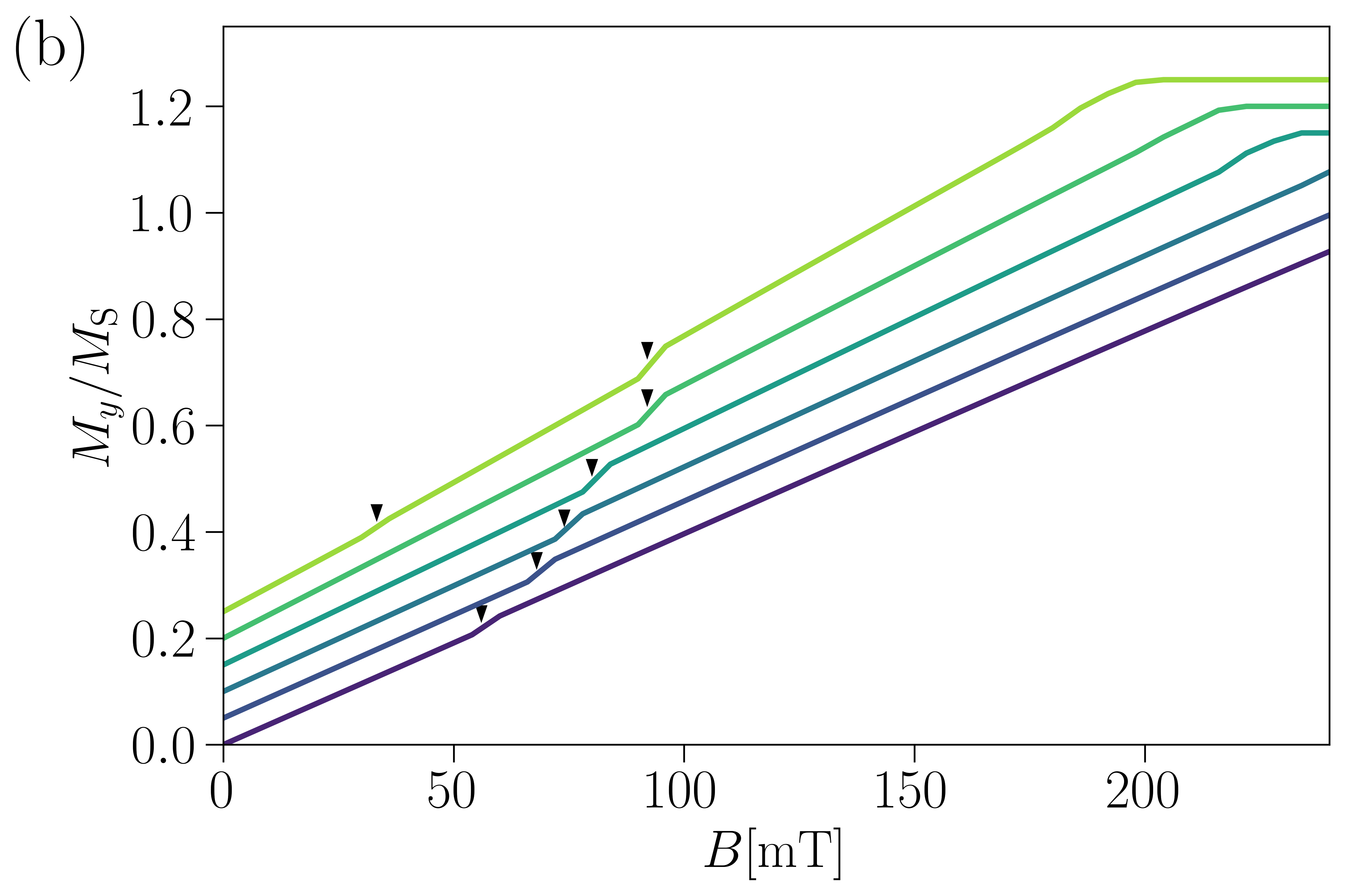}
}
\caption{
The equilibrium properties of the system: (a) the periodicity $L(B)$ and (b) the net magnetization along the magnetic field $M_{y}/M_{\mathrm{S}}$ for the equilibrium state as a function of the magnetic field. 
From blue to green the thicknesses of the systems are $t=36$, 60, 96, 156, 240, 348 nm. 
For clarity, in (b) we translate up the curves in steps of size $0.05$ and the black triangles mark the small jumps in the magnetization.
}
\label{fig:cort_diag_eq}
\end{figure}

The previous discussion shows that the differences between the three L-TEM intensity patterns are rather subtle and may lay out of the resolution capability of the used imaging technique.
This limitation is particularly important to distinguish the FAN and the STR states, since both have roughly similar periodicities. If a MFM imaging method is used instead, both states could be unambiguously identified as suggested by the first rows in Fig.~\ref{fig:config}.


The transitions between the different modulated states can be inferred from the behavior of $L(B)$ for the equilibrium state and the net magnetization along the magnetic field $M_{y}$. We show in Fig.~\ref{fig:cort_diag_eq} those magnitudes for different values of the thickness 36 nm$\leq t\leq 348$ nm (from blue to green), corresponding to different slides of the phase diagram in Fig.~\ref{fig:diag_eq}.
The transition from the CSL to the STR state is better observed in the behavior of $L(B)$, as it exhibits a discontinuity at the transition point. The value of $L(B)$ jumps to higher values with a step of height $\approx200$ nm. The transition from the STR to the FAN states is signaled by the presence of a peak in $L(B)$. For $B$ below the transition field the system is in the STR state and $L(B)$ grows with the magnetic field. For magnetic field above the transition field the state corresponds to the FAN state and the period decreases with the magnetic field.
The net magnetization along the magnetic field exhibits a tiny jump for the transition from the CSL to the STR state. A small jump is also observed when the system goes from the STR to the FAN state. Therefore, the identification of the magnetic transitions should be better resolved from magnetic susceptibility measurements or, as suggested by our previous discussion about the behavior of $L(B)$, from the measurement of the chacarcteristic length of the magnetic state as a function of the applied magnetic field. 

%
\subsection{The metastable states}
\label{sec:meta}

\begin{figure*}[htb]
\subfigure{
\includegraphics[width=5.5cm]{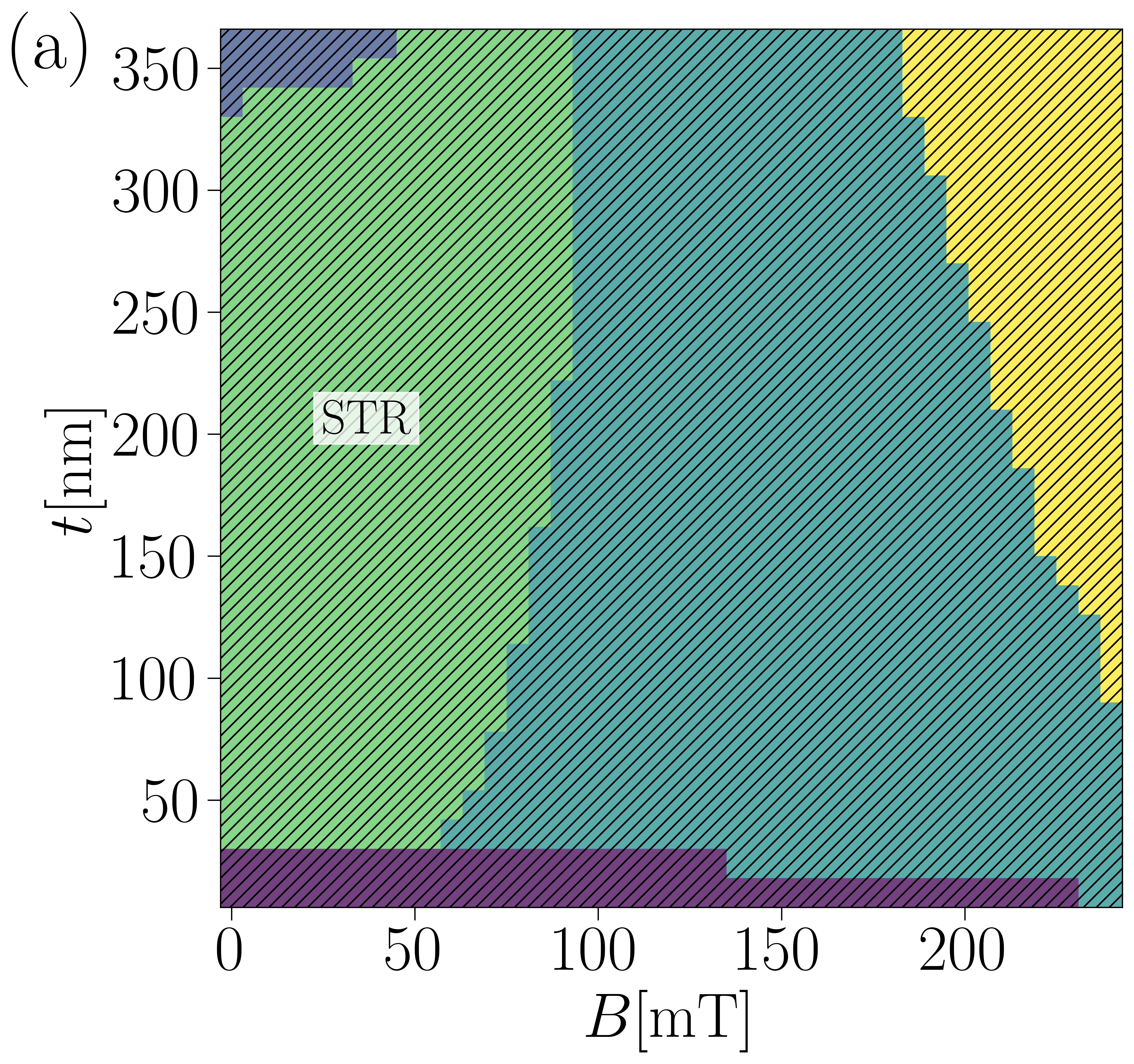}
}
\subfigure{
\includegraphics[width=5.5cm]{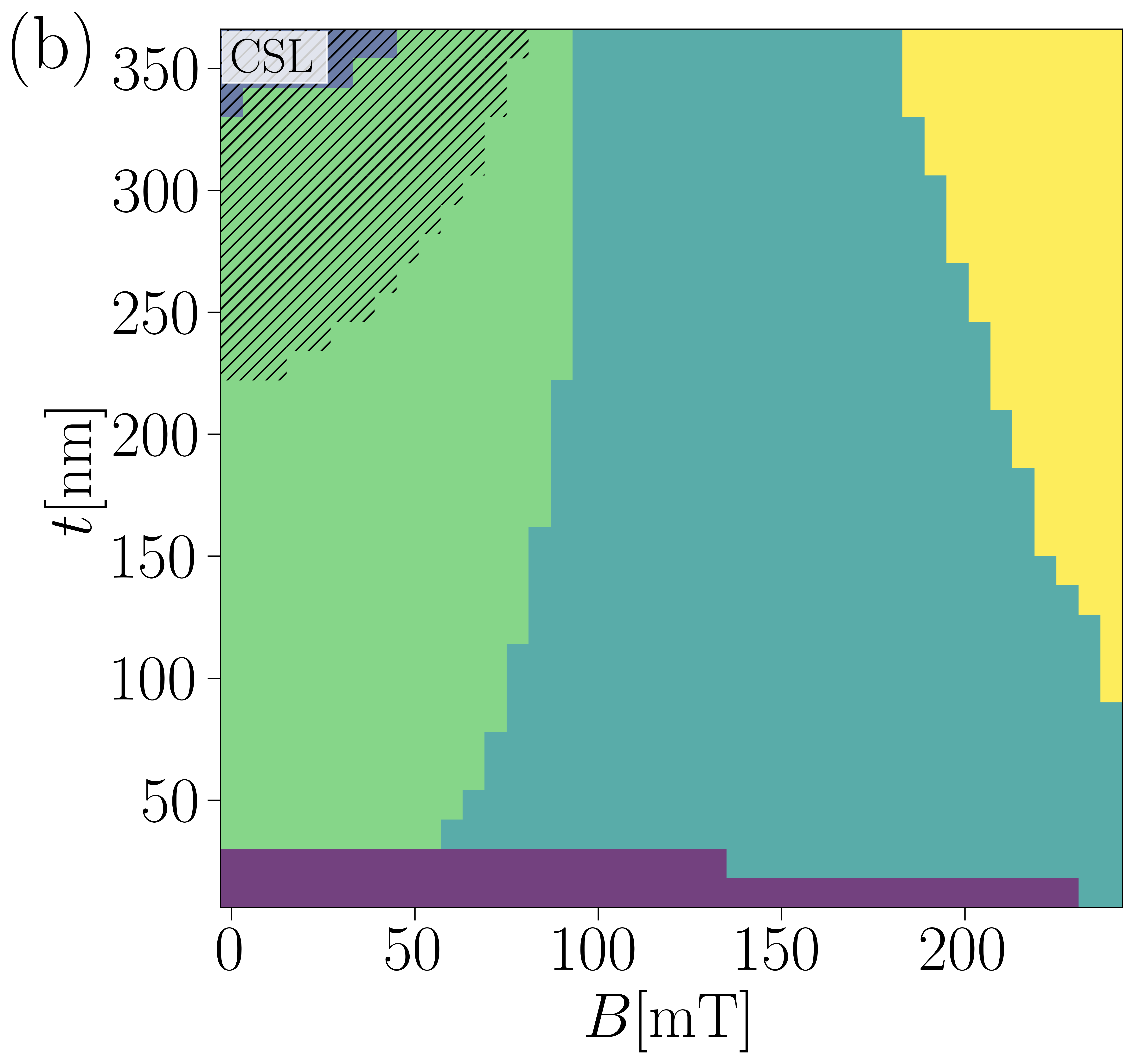}
}
\subfigure{
\includegraphics[width=5.5cm]{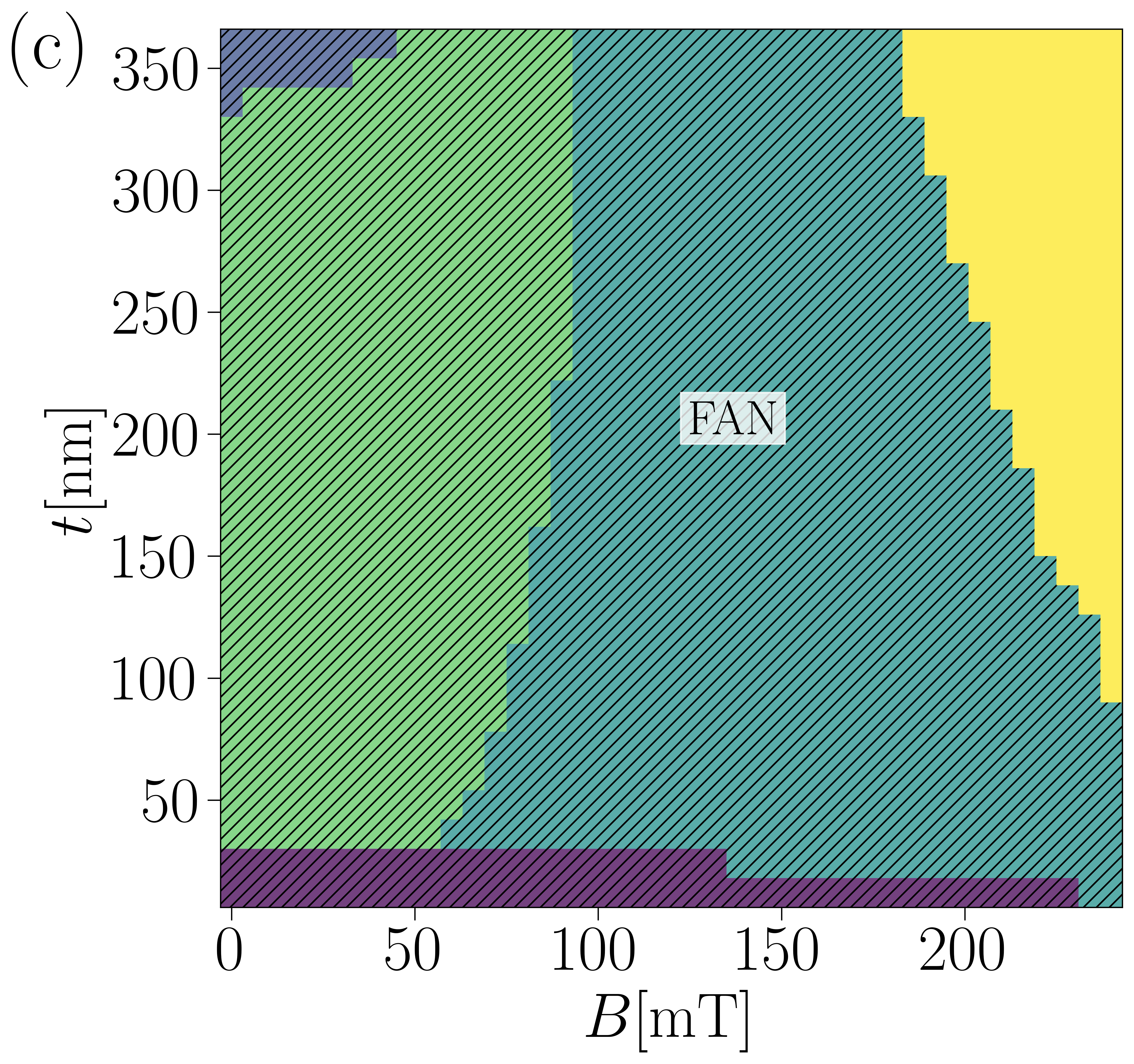}
}

\caption{
Metastability phase diagrams for: (a) the STR state, (b) the CSL state and (c) the FAN state. In each case, the hatched region represents the metastability of the magnetic states. As a reference, the colored background is the equilibrium phase diagram in Fig. ~\ref{fig:diag_eq}.
}
\label{fig:diag_meta}
\end{figure*}

Besides the equilibrium states, in some regions of the phase diagram there are metastable states determined by the presence of a local minimum in energy that does not necessarily coincide with the absolute minimum.
For instance, the energy curve obtained by relaxation from a HEC state presents a local minima (except at large enough magnetic fields) even if this state is not the equilibrium state of the system. 
The corresponding metastable state is a FAN state.
Analogously, the energy curve obtained by relaxation from a HOC state presents sometimes  two local minima, one around $L=200$ nm and another one above $L=500$ nm. They correspond to a CSL state and a STR state, respectively, and are metastable states (one of them may be the equilibrium state).
The first minimum (the CSL state) disappears in some regions of the phase diagram.
In Fig.~\ref{fig:diag_meta} we show the metastability region, represented by the hatched area, for the STR (Fig.~\ref{fig:diag_meta} (a)), the CSL (Fig.~\ref{fig:diag_meta} (b)) and the FAN (Fig.~\ref{fig:diag_meta} (c)) states. We observe that each phase can exist as a metastable state beyond the region where it is the equilibrium state. In Fig.~\ref{fig:diag_meta} (a) the metastability range of the STR state extends over the entire studied phase diagram. In Fig.~\ref{fig:diag_meta} (b) the metastability limit for the CSL corresponds to the values of $t$ and $B$ at which the first minimum (around 200 nm) disappears in the energy density obtained from the HOC state. Finally, in Fig.~\ref{fig:diag_meta} (c) the metastability region of the FAN state occupies a large portion of the phase diagram. The mestastabilty boundary coincides with the stability boundary of the FMY state. This is due to the fact that the FAN state gradually evolves to the FMY state as the magnetic field increases.

%


%
\subsection{Properties of metastable states}
\label{sec:periods}

\begin{figure*}[htb!]
\includegraphics[width=5.5cm]{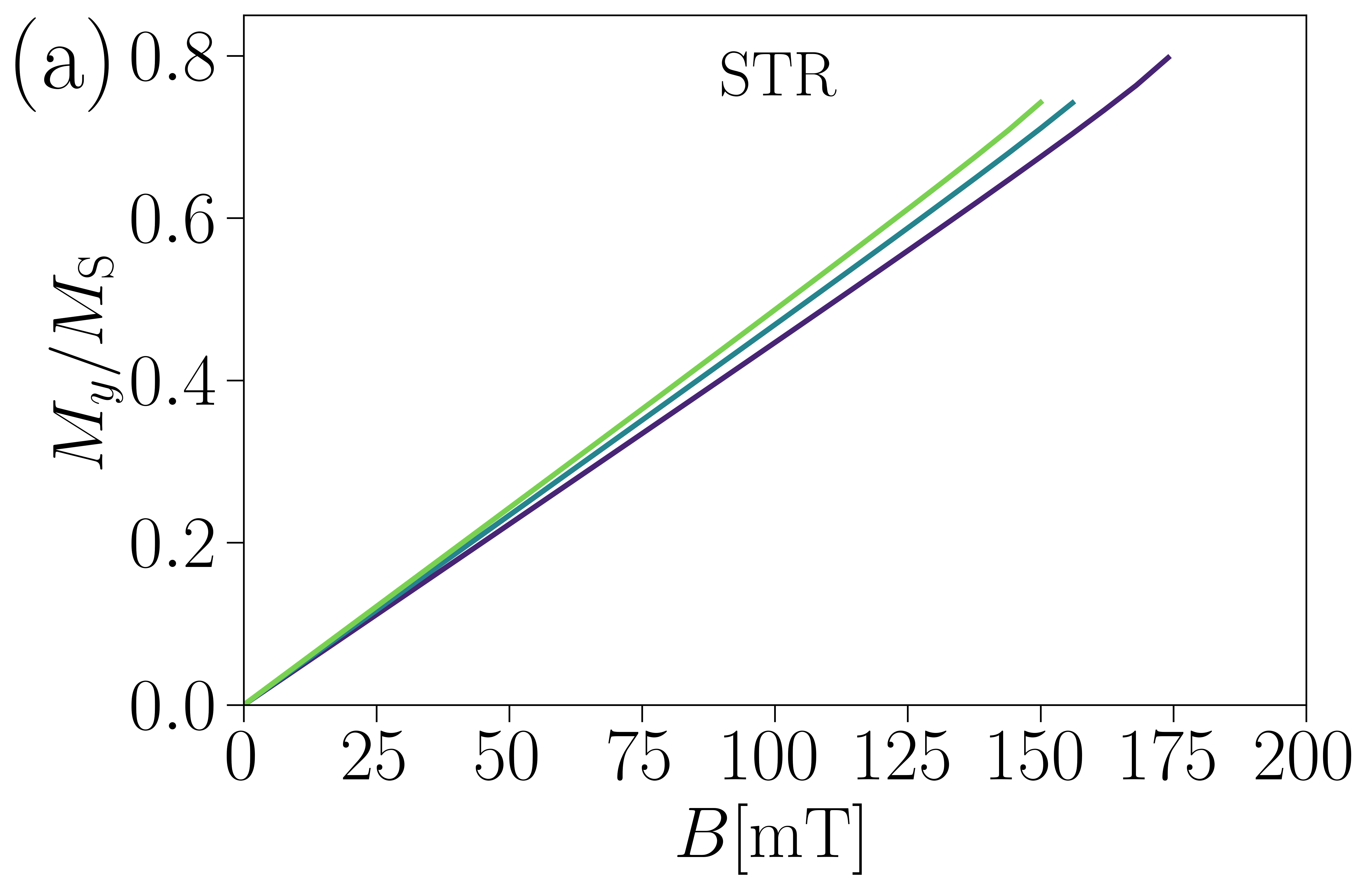}
\includegraphics[width=5.5cm]{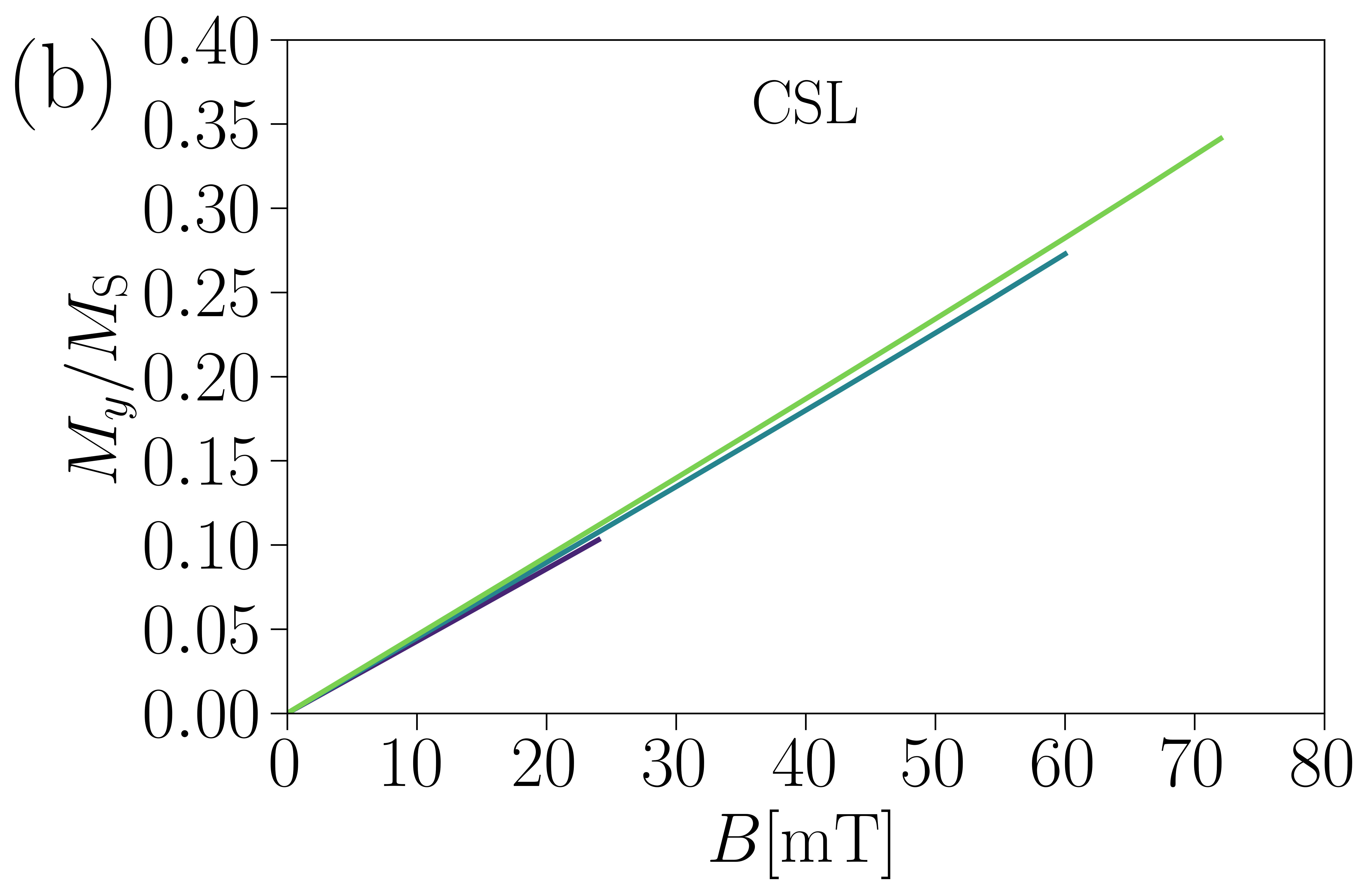}
\includegraphics[width=5.5cm]{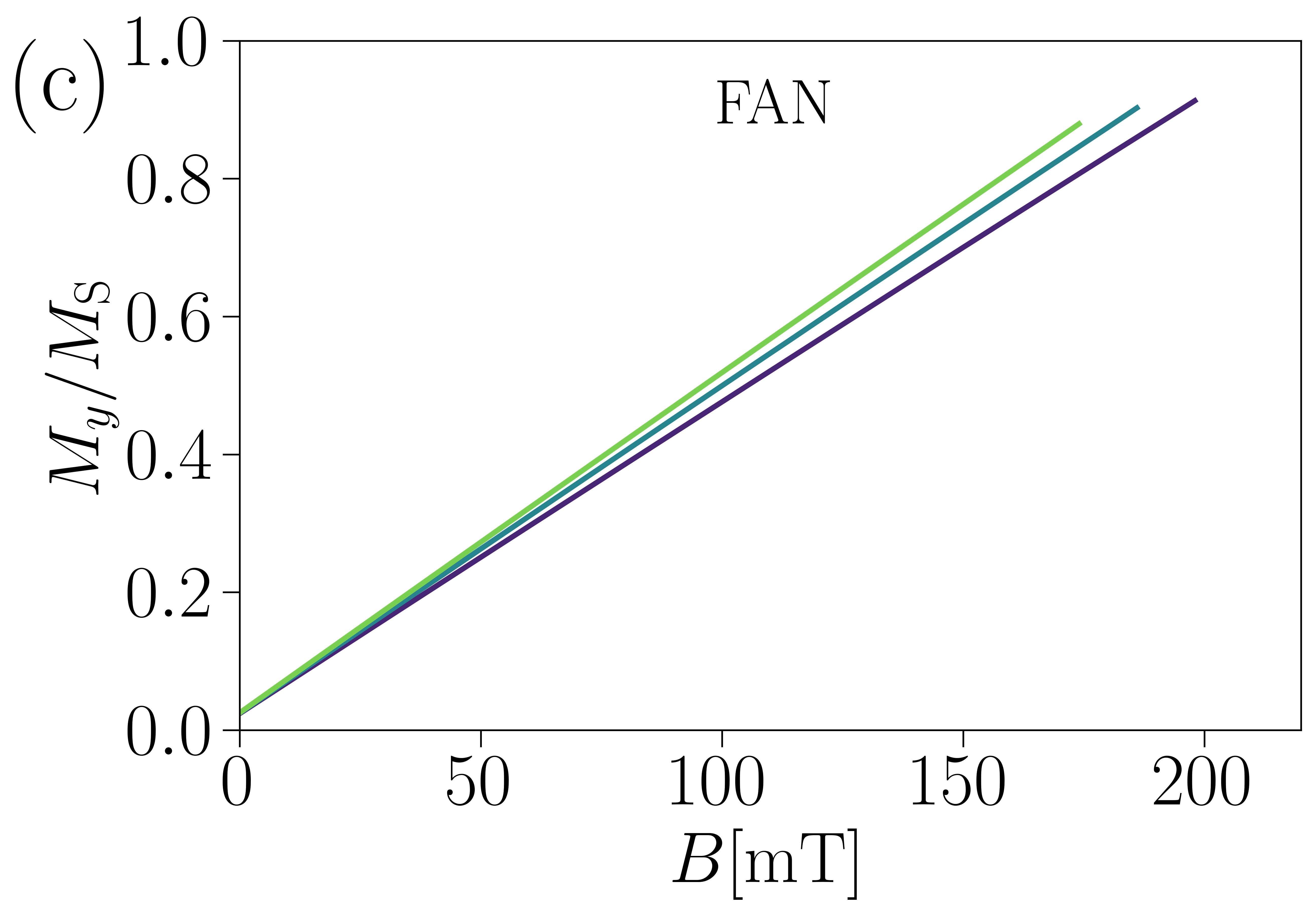}
\includegraphics[width=5.5cm]{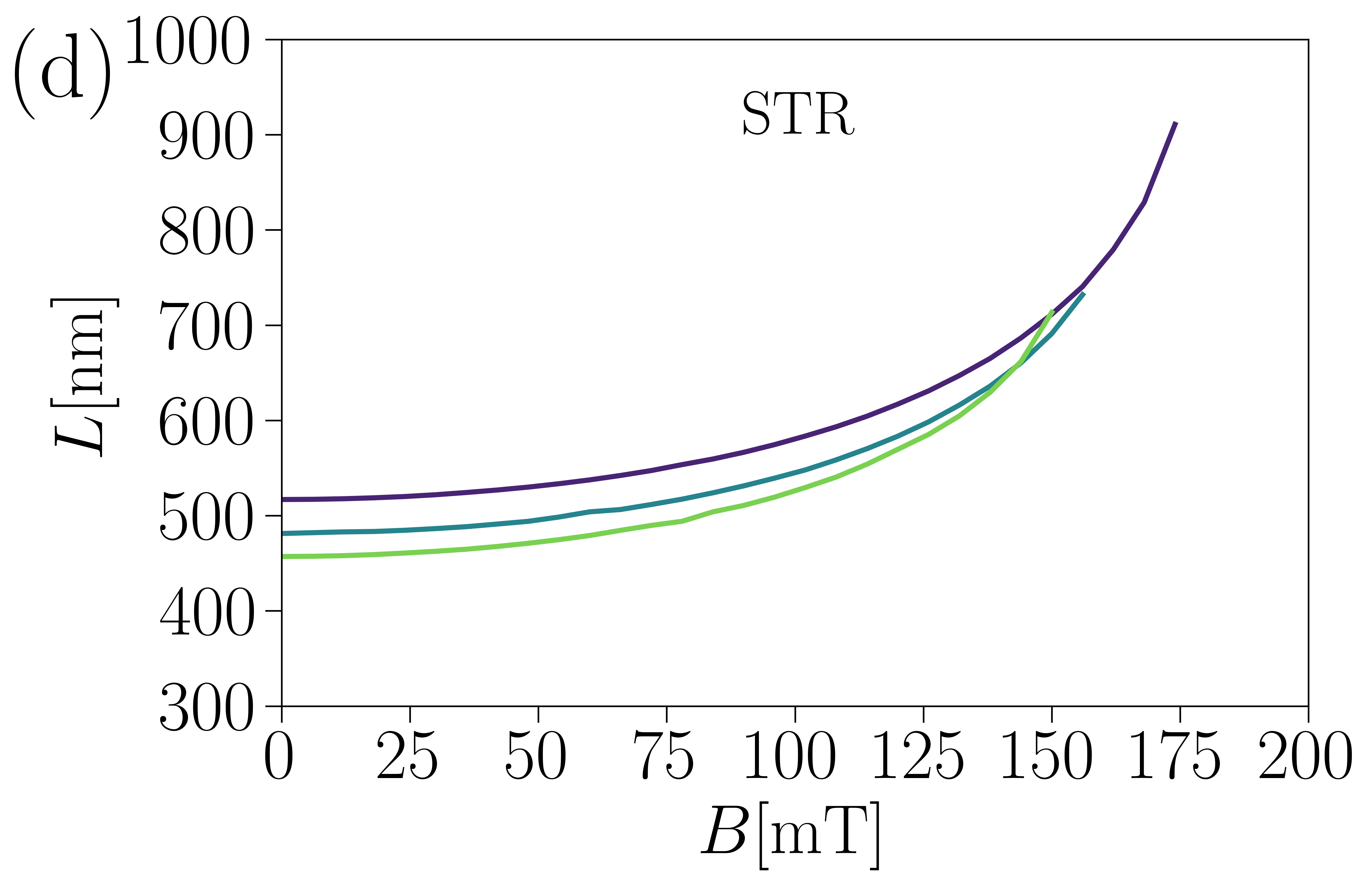}
\includegraphics[width=5.5cm]{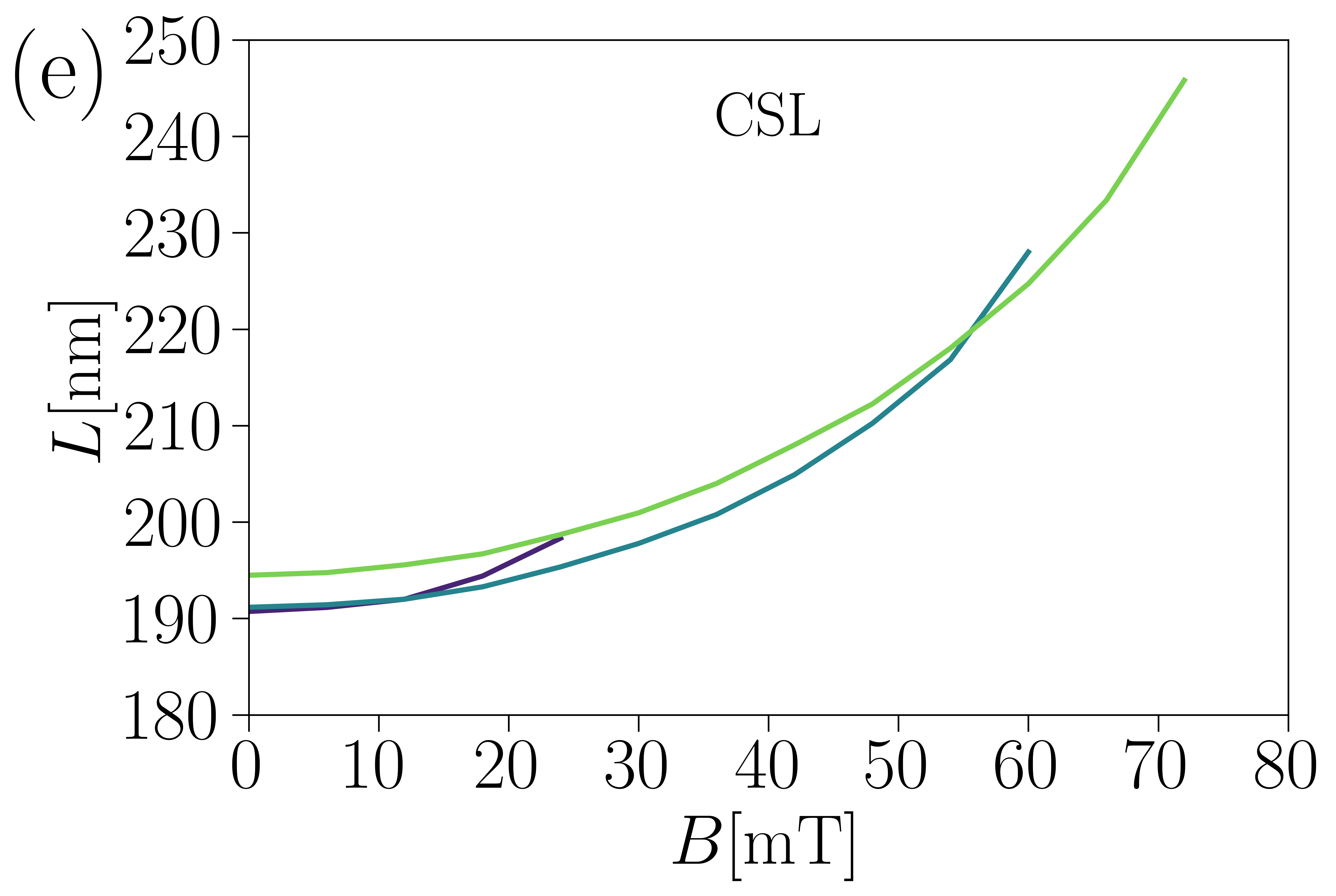}
\includegraphics[width=5.5cm]{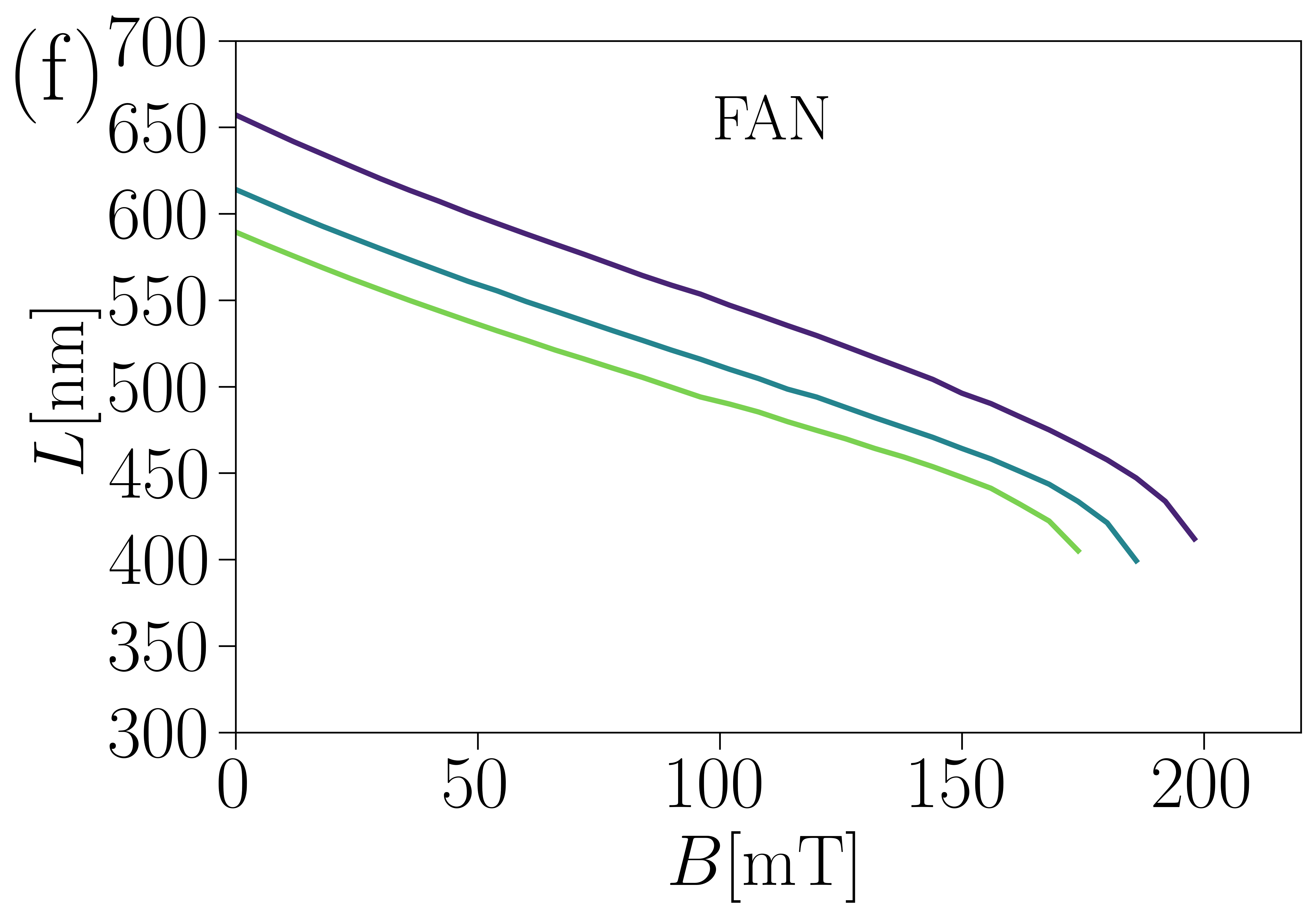}
\caption{Variation of $M_{y}/M_{\mathrm{S}}$ in (a)-(c) and $L$ in (d)-(f) as a function of $B$ for the STR, CSL and FAN states in the metastability region. The different colors represent different values of the thickness $t=240$ nm (in blue), $t=300$ nm (in watery green) and $t=360$ nm (in green).
}
\label{fig:curvas_meta}
\end{figure*}

Because of the metastability of the observed states, the system can be trapped in one of those states even if it does not correspond to the low energy state. We study here their properties within their metastability region. In particular we study the behaviour of the magnetization along the magnetic field and the period of the magnetic texture (Fig. \ref{fig:curvas_meta}).

In Fig. \ref{fig:curvas_meta} (a)-(c) we show the net magnetization along the magnetic field, which grows linearly with $B$ for the three modulated states. Except for their slope, the curves do not differ significantly, making it difficult to distinguish each metastable state only using magnetization curves.
However, one key feature to distinguish the modulated states from each other is the behavior of the periodicity of the state with the magnetic field and the thickness of the sample. This is particularly relevant to distinguish between the CSL and STR phases.

%

The period of the topologically non-trivial states (CSL and STR) increases with the magnetic field $B$ (Fig. \ref{fig:curvas_meta} (d)-(e)), while the period of the topologically trivial state (FAN) decreases with $B$ (Fig. \ref{fig:curvas_meta} (f)).
As we showed in the previous sections, the transition from the STR state to the FAN state occurs when the period of the STR state coincides with that of the FAN state. The periods of both states take values within roughly the same range, however they present opposite behaviors with the magnetic field even when they are metastable. These features are relevant to distinguish each state through their response to external magnetic fields.

\begin{figure}[htb!]
\includegraphics[width=8cm]{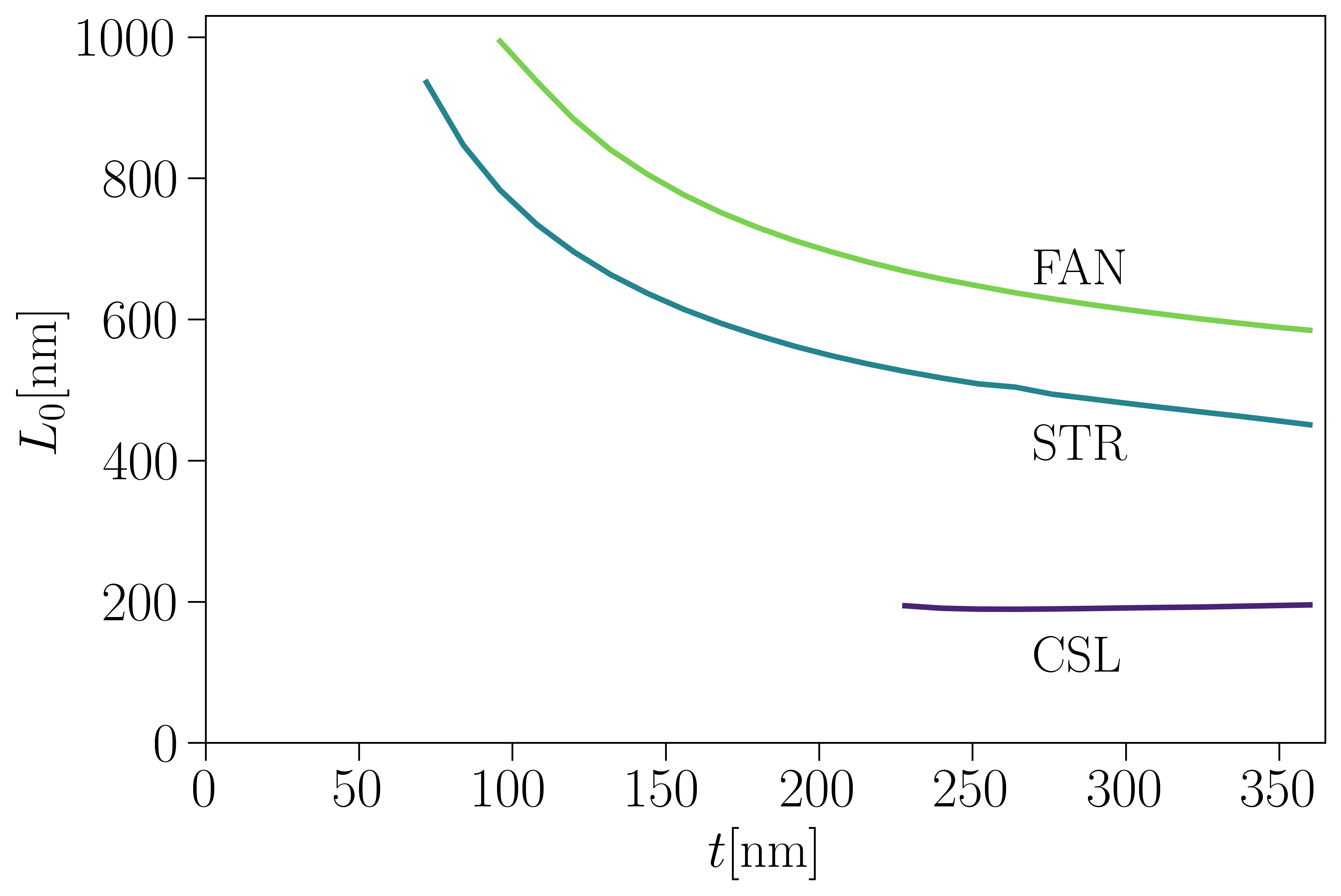}
\caption{Period at zero magnetic field ($L_{0}$) as a function of $t$ for the CSL (in blue), the STR (in watery green) and the FAN (in green). 
}
\label{fig:l0_es}
\end{figure}

Besides the factor $\sim 2$ in the typical length scales that characterize the CSL and the STR states, there is another important feature that distinguishes both states. As shown in Fig.~\ref{fig:l0_es}, the value of the periodicity at $B=0$ ($L_{0}=L(0)$), strongly depends on the thickness of the system and decreases as $t$ grows for the STR state. On the contrary, $L_{0}$ shows a neglectable dependence on $t$ for the CSL. This reinforces the idea that the  typical length scale and the main properties of the CSL can be mostly attributed to the DMI.
The value of $L_{0}$ for the STR and FAN states decrease with increasing values of the thickness of the sample as shown in Fig. ~\ref{fig:l0_es}. Thus the dipolar interaction play a relevant role in the properties of these states.
Finally, it is important to observe that the value of $L_{0,\mathrm{CSL}}$ is smaller than the value expected for the system if the dipolar field were absent, $L_{0}=4\pi A/D\approx250$ nm ($L_{0,\mathrm{CSL}}\approx 0.8 L_{0}$). This implies that the dipolar field affects the period of the CSL structure.

\section{The effect of the DMI}
\label{sec:dmi_effect}

The details on the sample preparation could lead to slight variations in the DMI constant. The results discussed in the previous sections suggest that the key properties of these kind of magnets are determined mainly by the balance between the DMI and the dipolar interaction, and therefore they may exhibit a strong dependence on the value of the DMI strength.  
Therefore, it is interesting to analyze the effect of a variation in the value of $D$. To this end we let the DMI constant vary within the range, $0.9D_{0}\leq D\leq1.1D_{0}$, around the reference value $D_{0}=50$ $\mu$J$/$m$^{2}$ for the MnNb$_{3}$S$_{6}$ compound. In particular we study the properties of the system at $B=0$ mT for different thicknesses $t=156$, 252, 348 nm.

\begin{figure}[p]
\subfigure{
\includegraphics[width=8cm]{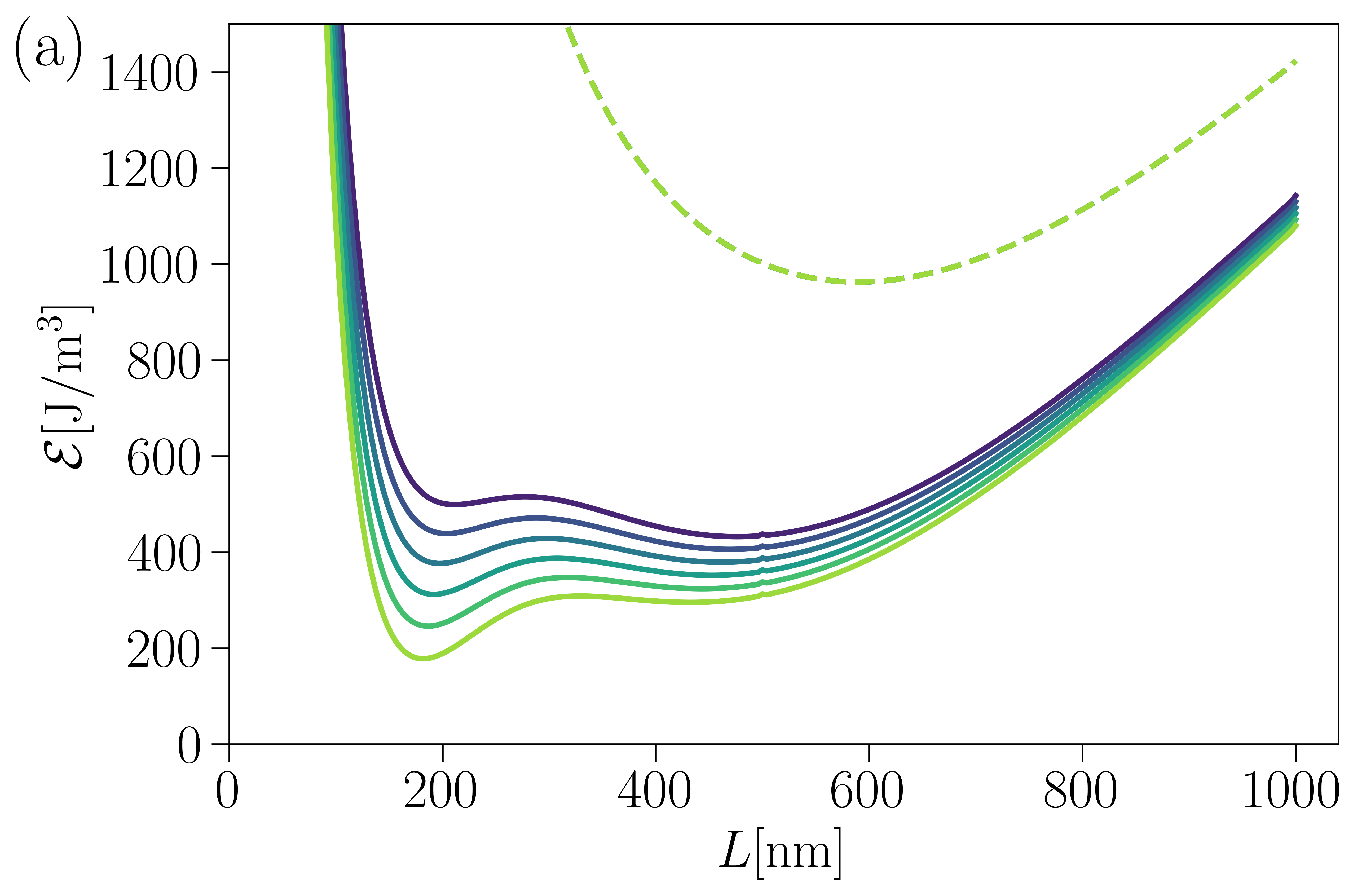}
}
\subfigure{
\includegraphics[width=8cm]{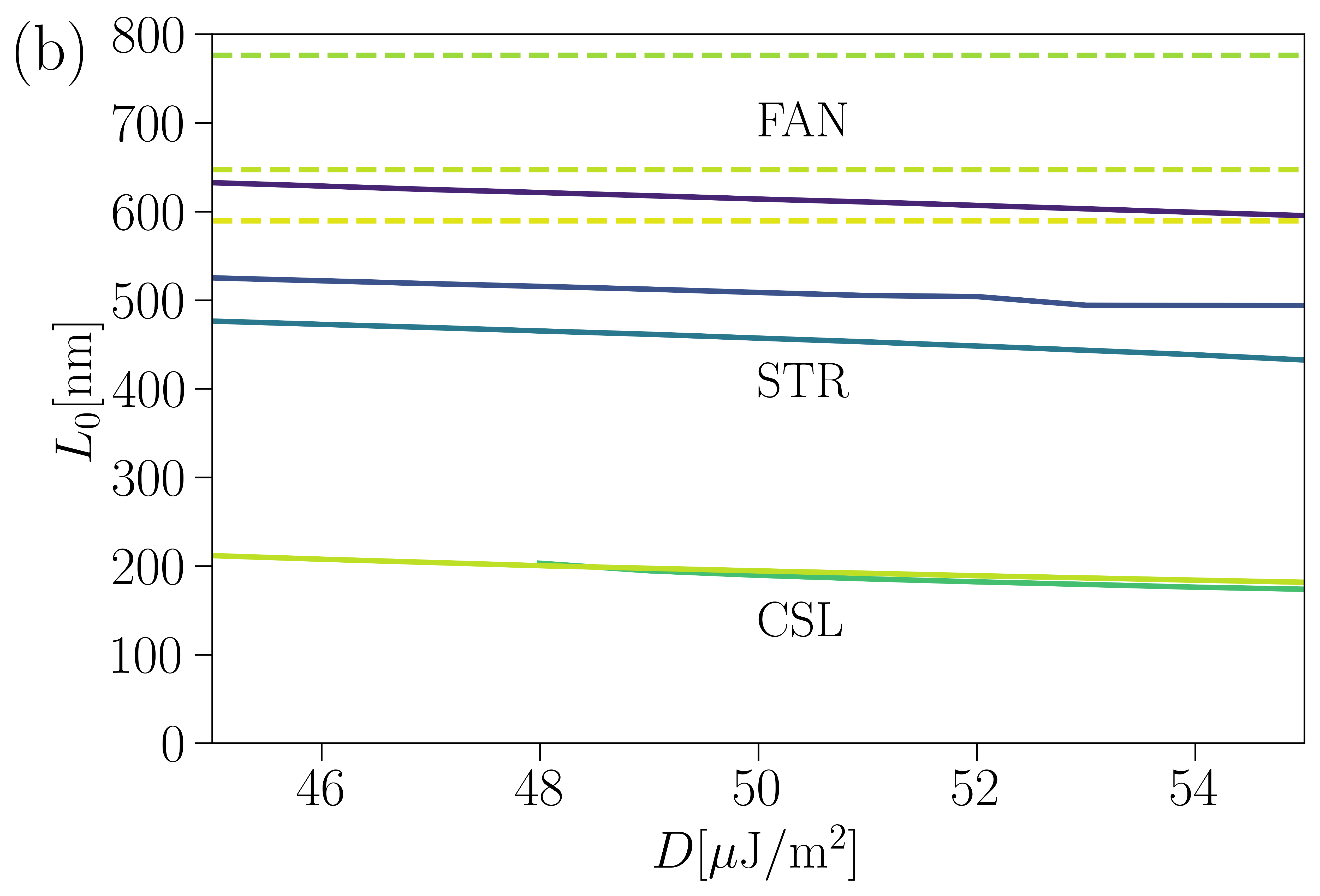}
}
\subfigure{
\includegraphics[width=8cm]{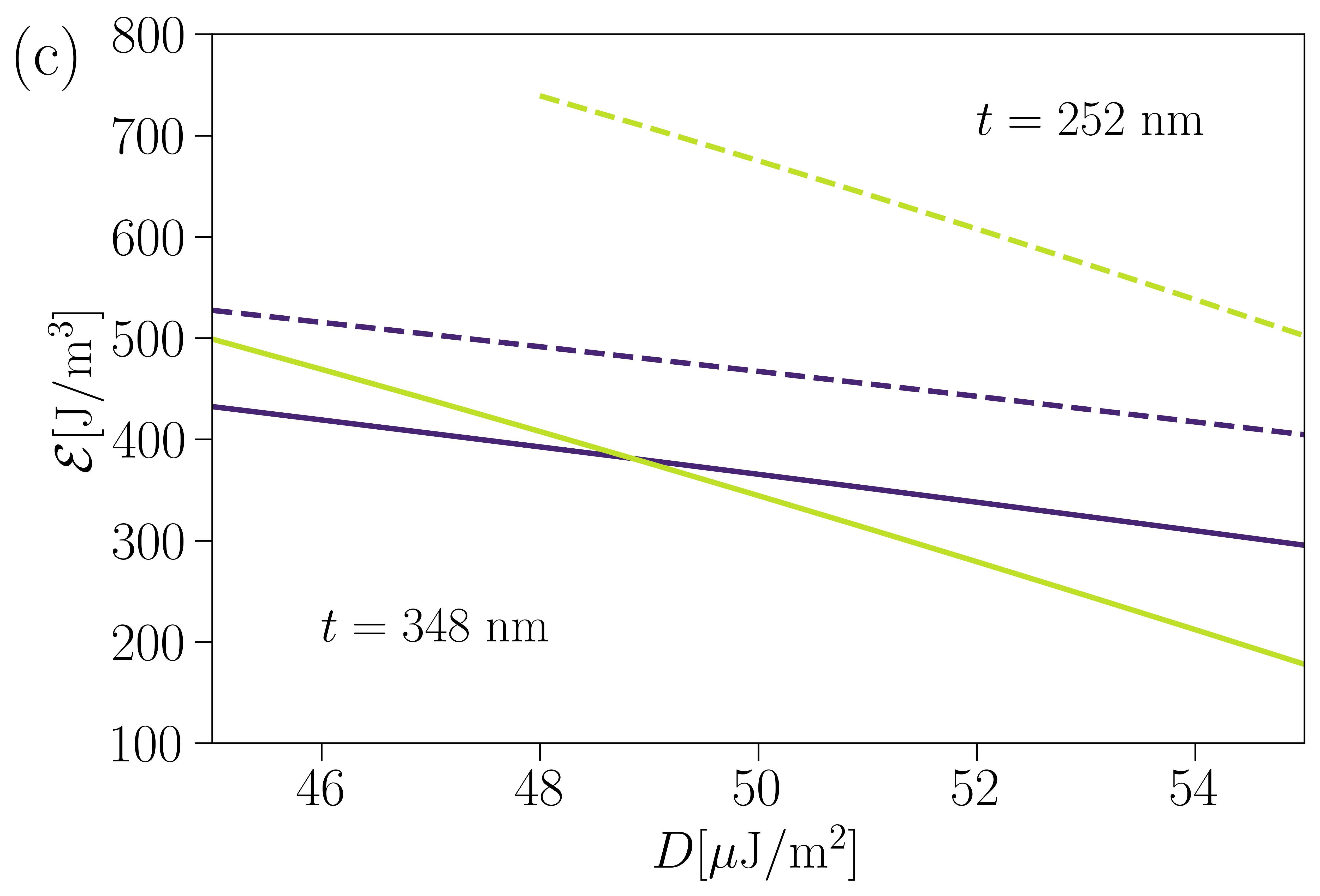}
}
\caption{
(a) Energy density as a function of the period of the magnetic texture for $B=0$ mT and $t=348$ nm for different values of the DMI constant for the state that results after a relaxation from the initial condition corresponding to the HOC configuration (solid lines) and to the HEC (dashed lines). The value of $D$ varies between $0.9 D_{0}$ (in blue) and $1.1 D_{0}$ (in green), with $D_0 = 50$~$\mu$J/m$^2$. All the curves obtained from the HEC state (dashed lines) are overlapping.
(b) Dependence of the period with $D$, for the states that results after a relaxation from the initial condition corresponding to the HOC configuration (solid lines) and the HEC (dashed lines). The relaxation of the HOC initial configuration can lead to a STR (in blue) or a CSL (in yellow-green).
(c) Variation of the energy density as a function of $D$ for the CSL (in yellow-green) and the STR (in blue) states at two different thicknesses: $t=348$ nm (solid lines) and $t=252$ nm (dashed lines).
}
\label{fig:land_dmi}
\end{figure}

The energy density curve for the state resulting from the HEC initial condition (that relaxes to the FAN state) is not affected by the value of the DMI coupling as shown in Fig.~\ref{fig:land_dmi} (a) (in dashed lines, which constitutes a set of overlapping curves). Therefore, the periodicity of this modulated state does not depend on the value of the DMI constant as shown in Fig. \ref{fig:land_dmi} (b) (dashed lines).
The authors in Ref. \onlinecite{Karna2021} suggest that the interactions between the solitons and antisolitons could be mediated by the DMI. However our results indicate that this may not be the case for solitons with opposite chiralities, as it is the case of the FAN state. The main features of the FAN state rely in the remaining interactions.
On the contrary, the energy density of the state resulting from the HOC initial state, exhibits a strong dependence on the value of the DMI coupling as shown in Fig. \ref{fig:land_dmi} (a) for $t=348$ nm (solid lines). Thus the interaction between solitons with the same chirality could be mediated by the DMI. One important effect is that the energy density in the first minimum (around 200 nm) is very sensitive to the value of $D$, and this minimum can change from a local minimum to a global one as the DMI constant grows. As a consequence, $L_{0}$ decreases for the STR and the CSL with increasing values of $D$ as shown in Fig.~\ref{fig:land_dmi} (b) (solid lines), which differ from the results for the FAN state.
Finally, we observe in Fig.~\ref{fig:land_dmi} (c) that the energy density of the CSL (in green) exhibits a stronger dependence with $D$ than that of the STR (in blue).
It is important to note from this figure that a change in $D$ of $\approx 10$ \% can have an important impact in the structure of the phase diagram. For instance at $t=348$ nm (and $B=0$ mT) the ground state changes from STR to CSL and the transition is at $D \approx 49$ $\mu$J/m$^{2}$. 
Another important result from this figure is that, at $B=0$ mT, for thinner systems a larger value of $D$ is required to have the CSL as a ground state. As shown in Fig.~\ref{fig:land_dmi} (c), for the system with $t=252$ nm a large value of the DMI (outside the analyzed range) is needed to find the system in the CSL as a ground state. These effects are also a manifestation of the interplay between the dipolar field and the DMI in the determination of the magnetic properties of the system and in particular its phase diagram.

\section{Summary and Discussion}
\label{sec:conc}
%

We have shown that the presence of a strong dipolar interaction, as compared with the DMI, lead to a series of states which are close in energy. In particular the presence of a second local minimum in the energy density of the HOC state, which is not observed when just one interaction (either the DMI or the dipolar interaction) is taken into account, could be considered as a criterion to decide whether both interactions are relevant.
The complexity in understanding the magnetic properties of compounds such as MnNb$_3$S$_6$ has its origin in the competition of three different modulated states: the CSL, the STR and the FAN states. We have studied and characterized the magnetic structures of each of those states, and how the ground state of the system changes with its thickness and the magnetic field. We have also shown that the three modulated states can be found beyond their equilibrium region as metastable states.


Our results present a challenging scenario from the experimental point of view, explaining why some experimental findings continue puzzling the scientific community.
For instance, we have shown that the CSL is distorted due to the dipolar interaction and that on the surface of the sample this state might look like a series of ferromagnetic domains connected by usual domain walls. We believe that the distorted helical magnetic structure observed in Ref. \onlinecite{Karna2021} could be probably interpreted as the CSL in our simulations. However the presence of surface domain-like patterns may hinder the helical structure of the state in absorption and AC susceptibility measurements \cite{dai2020microwave,hall2022comparative}. 
Along a similar line, indirect measurements suggest that $L(B)/L_{0}$ has a similar behaviour in MnNb$_3$S$_6$ and in CrNb$_3$S$_6$ \cite{li2023temperature}. However we stress that this behaviour is roughly shared by the CSL and the STR as shown in Fig. \ref{fig:curvas_meta} (d) and (e). Thus, this method does not unambiguously identifies the observed state.

Regarding the magnetic transitions, there is a great dispersion in the reported values for the critical fields associated to the transition from a helical order to the forced ferromagnetic order and there are also additional transitions that have not yet been fully characterized \cite{kousaka2009chiral,dai2019critical, karna2019consequences, Karna2021,hall2022comparative, ohkuma2022nonequilibrium}.
Except for the values of the critical field reported in Refs. \onlinecite{kousaka2009chiral, ohkuma2022nonequilibrium} (in which the critical field seems to be much smaller than in the rest of the literature) the critical fields are roughly within the values that we encountered in our simulations for thin systems. Now, if we neglect the dipolar field in Eq. \eqref{eq:model}, to consider the bulk limit, and compute the critical field for the transition from the CSL to the field polarized state  $B_{c}=\frac{\pi^{2}}{32}\frac{D^{2}}{M_{\mathrm{S}}A}$ \cite{izyumov1984modulated, kishine2005synthesis, Bostrem08b, Kishine09, Kishine10, Kishine11, togawa2012chiral, victor2020dynamics}, we obtain $B_{c}\approx3.59$ mT. Remarkably, this value of $B_{c}$ is consistent with the one found by the authors in Ref. \onlinecite{kousaka2009chiral} for a magnetic transition interpreted as the transition from the CSL to the uniform ferromagnetic state. In their experiments in large samples, with a thickness $t = 100$~$\mu$m well above the values considered in our work, the dipolar interaction may not play a relevant role or may contribute with an effective anisotropy term. 
Thus the complete model, as considered in our work, constitutes the basic core to interpret the phenomenology of this kind of magnets, from their bulk to their thin film properties.

Despite of the potential of the complete model of Eq. \eqref{eq:model} to explain the phenomenology of the compound MnNb$_3$S$_6$, there are some known facts that were left out of our analysis and may be related to additional ingredients such as thermal and disorder effects.
For instance, although in our model the FAN state appears as a ground state only for $B\gtrsim 90$ mT, in Ref. \cite{Karna2021}
the authors experimentally observe a state that can be generally interpreted as the FAN state for $B=0$ mT.
This could be the case when thermal fluctuations (not included in our model) enhance the stability of this state even at low magnetic fields~\cite{shinozaki2017fan}. Thermal fluctuations could also lead to heterogeneous magnetization field, with regions or domains mixing the FAN and STR states \cite{Karna2021}.
Recent research on the MnNb$_3$S$_6$ compound show that for temperatures below $T\approx15$ K the critical fields exponentially increase as the temperature decreases \cite{dai2019critical,ohkuma2022nonequilibrium}. This effect, which is absent in CrNb$_3$S$_6$ \cite{Togawa13,Ghimire13,Tsuruta16} and may lead to modifications in the parameters or the structure of the model in Eq. \eqref{eq:model},  was not taken into account in our analysis.
Finally, it was reported the lack of strict periodicity of the modulated states \cite{Karna2021, hall2022comparative}. This can be attributed to the presence of defects and pining effects which could hinder the strict periodicity of the periodic phases discussed here \cite{Karna2021,hall2022comparative}. These effects may also affect the structure of the phase diagram or even the stability of the modulated states, turning some of them into unstable states.

Our work shed light in the study of the magnetic properties of monoaxial chiral helimagnets, such as MnNb$_3$S$_6$, in which both the DMI as well as the dipolar interaction need to be taken into account to fully understand their properties. From a helical order \cite{kousaka2009chiral, karna2019consequences, dai2019critical, dai2020microwave, Karna2021, ohkuma2022nonequilibrium, li2023temperature} to ferromagnetic domains separated by conventional domain walls \cite{karna2019consequences, hall2022comparative}, different interpretations for the underlying magnetic states in MnNb$_3$S$_6$ have been suggested. 
Our work provides a scenario that can help unifying earlier and seemingly antagonistic interpretations, so that all those facts may be different facets of the interplay between the DMI and the dipolar interaction.

Besides the complex equilibrium properties, the presence of metastable states represents an additional challenge in the determination of the magnetic properties of chiral heliamgnets with strong dipolar effects. The three modulated states (the CSL, STR and FAN) can have similar properties as we saw for the case of the L-TEM images and the magnetization curves. So a given technique may fail to spot the signature of each state. For these reasons, further experimental research, that in particular include complementary techniques, are necessary to fully understand the physics of compounds like MnNb$_3$S$_6$. Also further theoretical research is needed to identify the fingerprints that each magnetic state can present in different experiments. In particular, the simulation of ferromagnetic resonance experiments and AC magnetic susceptibility experiments could help to identify the different modulated states.

\begin{acknowledgments}
Grant No. PID2022-138492NB-I00-XM4 funded by MCIN/AEI/10.13039/501100011033 supported this work. Grants OTR02223-SpINS from CSIC/MICIN and DGA/M4 from Diputaci\'on General de Arag\'on (Spain) are also acknowledged. This work was also supported by the Grant No. PICT 2017-0906 from the Agencia Nacional de Promoci\'on Cient\'ifica y Tecnol\'ogica, Argentina.
\end{acknowledgments}









\bibliography{references}

\begin{thebibliography}{49}%
\makeatletter
\providecommand \@ifxundefined [1]{%
 \@ifx{#1\undefined}
}%
\providecommand \@ifnum [1]{%
 \ifnum #1\expandafter \@firstoftwo
 \else \expandafter \@secondoftwo
 \fi
}%
\providecommand \@ifx [1]{%
 \ifx #1\expandafter \@firstoftwo
 \else \expandafter \@secondoftwo
 \fi
}%
\providecommand \natexlab [1]{#1}%
\providecommand \enquote  [1]{``#1''}%
\providecommand \bibnamefont  [1]{#1}%
\providecommand \bibfnamefont [1]{#1}%
\providecommand \citenamefont [1]{#1}%
\providecommand \href@noop [0]{\@secondoftwo}%
\providecommand \href [0]{\begingroup \@sanitize@url \@href}%
\providecommand \@href[1]{\@@startlink{#1}\@@href}%
\providecommand \@@href[1]{\endgroup#1\@@endlink}%
\providecommand \@sanitize@url [0]{\catcode `\\12\catcode `\$12\catcode
  `\&12\catcode `\#12\catcode `\^12\catcode `\_12\catcode `\%12\relax}%
\providecommand \@@startlink[1]{}%
\providecommand \@@endlink[0]{}%
\providecommand \url  [0]{\begingroup\@sanitize@url \@url }%
\providecommand \@url [1]{\endgroup\@href {#1}{\urlprefix }}%
\providecommand \urlprefix  [0]{URL }%
\providecommand \Eprint [0]{\href }%
\providecommand \doibase [0]{https://doi.org/}%
\providecommand \selectlanguage [0]{\@gobble}%
\providecommand \bibinfo  [0]{\@secondoftwo}%
\providecommand \bibfield  [0]{\@secondoftwo}%
\providecommand \translation [1]{[#1]}%
\providecommand \BibitemOpen [0]{}%
\providecommand \bibitemStop [0]{}%
\providecommand \bibitemNoStop [0]{.\EOS\space}%
\providecommand \EOS [0]{\spacefactor3000\relax}%
\providecommand \BibitemShut  [1]{\csname bibitem#1\endcsname}%
\let\auto@bib@innerbib\@empty
\bibitem [{\citenamefont {Dzyaloshinskii}(1958)}]{Dzyal58}%
  \BibitemOpen
  \bibfield  {author} {\bibinfo {author} {\bibfnamefont {I.}~\bibnamefont
  {Dzyaloshinskii}},\ }\bibfield  {title} {\bibinfo {title} {A thermodynamic
  theory of “weak” ferromagnetism of antiferromagnetics},\ }\href@noop {}
  {\bibfield  {journal} {\bibinfo  {journal} {J. Phys. Chem. Solids}\ }\textbf
  {\bibinfo {volume} {4}},\ \bibinfo {pages} {241} (\bibinfo {year}
  {1958})}\BibitemShut {NoStop}%
\bibitem [{\citenamefont {Moriya}(1960)}]{moriya1960new}%
  \BibitemOpen
  \bibfield  {author} {\bibinfo {author} {\bibfnamefont {T.}~\bibnamefont
  {Moriya}},\ }\bibfield  {title} {\bibinfo {title} {New mechanism of
  anisotropic superexchange interaction},\ }\href@noop {} {\bibfield  {journal}
  {\bibinfo  {journal} {Phys. Rev. Lett.}\ }\textbf {\bibinfo {volume} {4}},\
  \bibinfo {pages} {228} (\bibinfo {year} {1960})}\BibitemShut {NoStop}%
\bibitem [{\citenamefont {Dzyaloshinskii}(1964)}]{Dzyal64}%
  \BibitemOpen
  \bibfield  {author} {\bibinfo {author} {\bibfnamefont {I.}~\bibnamefont
  {Dzyaloshinskii}},\ }\bibfield  {title} {\bibinfo {title} {Theory of
  helicoidal structures in antiferromagnets. {I. Nonmetals}},\ }\href@noop {}
  {\bibfield  {journal} {\bibinfo  {journal} {Sov. Phys. JETP}\ }\textbf
  {\bibinfo {volume} {19}},\ \bibinfo {pages} {960} (\bibinfo {year}
  {1964})}\BibitemShut {NoStop}%
\bibitem [{\citenamefont {Moriya}\ and\ \citenamefont
  {Miyadai}(1982)}]{Moriya82}%
  \BibitemOpen
  \bibfield  {author} {\bibinfo {author} {\bibfnamefont {T.}~\bibnamefont
  {Moriya}}\ and\ \bibinfo {author} {\bibfnamefont {T.}~\bibnamefont
  {Miyadai}},\ }\bibfield  {title} {\bibinfo {title} {Evidence for the helical
  spin structure due to antisymmetric exchange interaction in
  {Cr$_{1/3}$NbS$_2$}},\ }\href@noop {} {\bibfield  {journal} {\bibinfo
  {journal} {Solid State Commun.}\ }\textbf {\bibinfo {volume} {42}},\ \bibinfo
  {pages} {209} (\bibinfo {year} {1982})}\BibitemShut {NoStop}%
\bibitem [{\citenamefont {Miyadai}\ \emph {et~al.}(1983)\citenamefont
  {Miyadai}, \citenamefont {Kikuchi}, \citenamefont {Kondo}, \citenamefont
  {Sakka}, \citenamefont {Arai},\ and\ \citenamefont {Ishikawa}}]{Miyadai83}%
  \BibitemOpen
  \bibfield  {author} {\bibinfo {author} {\bibfnamefont {T.}~\bibnamefont
  {Miyadai}}, \bibinfo {author} {\bibfnamefont {K.}~\bibnamefont {Kikuchi}},
  \bibinfo {author} {\bibfnamefont {H.}~\bibnamefont {Kondo}}, \bibinfo
  {author} {\bibfnamefont {S.}~\bibnamefont {Sakka}}, \bibinfo {author}
  {\bibfnamefont {M.}~\bibnamefont {Arai}},\ and\ \bibinfo {author}
  {\bibfnamefont {Y.}~\bibnamefont {Ishikawa}},\ }\bibfield  {title} {\bibinfo
  {title} {Magnetic properties of {Cr$_{1/3}$NbS$_2$}},\ }\href@noop {}
  {\bibfield  {journal} {\bibinfo  {journal} {J. Phys. Soc. Jpn.}\ }\textbf
  {\bibinfo {volume} {52}},\ \bibinfo {pages} {1394} (\bibinfo {year}
  {1983})}\BibitemShut {NoStop}%
\bibitem [{\citenamefont {Izyumov}(1984)}]{izyumov1984modulated}%
  \BibitemOpen
  \bibfield  {author} {\bibinfo {author} {\bibfnamefont {Y.~A.}\ \bibnamefont
  {Izyumov}},\ }\bibfield  {title} {\bibinfo {title} {Modulated, or
  long-periodic, magnetic structures of crystals},\ }\href@noop {} {\bibfield
  {journal} {\bibinfo  {journal} {Sov. Phys. Usp.}\ }\textbf {\bibinfo {volume}
  {27}},\ \bibinfo {pages} {845} (\bibinfo {year} {1984})}\BibitemShut
  {NoStop}%
\bibitem [{\citenamefont {Kishine}\ \emph {et~al.}(2005)\citenamefont
  {Kishine}, \citenamefont {Inoue},\ and\ \citenamefont
  {Yoshida}}]{kishine2005synthesis}%
  \BibitemOpen
  \bibfield  {author} {\bibinfo {author} {\bibfnamefont {J.-i.}\ \bibnamefont
  {Kishine}}, \bibinfo {author} {\bibfnamefont {K.}~\bibnamefont {Inoue}},\
  and\ \bibinfo {author} {\bibfnamefont {Y.}~\bibnamefont {Yoshida}},\
  }\bibfield  {title} {\bibinfo {title} {Synthesis, structure and magnetic
  properties of chiral molecule-based magnets},\ }\href@noop {} {\bibfield
  {journal} {\bibinfo  {journal} {Prog. Theo. Phys. Supp.}\ }\textbf {\bibinfo
  {volume} {159}},\ \bibinfo {pages} {82} (\bibinfo {year} {2005})}\BibitemShut
  {NoStop}%
\bibitem [{\citenamefont {Togawa}\ \emph {et~al.}(2012)\citenamefont {Togawa},
  \citenamefont {Koyama}, \citenamefont {Takayanagi}, \citenamefont {Mori},
  \citenamefont {Kousaka}, \citenamefont {Akimitsu}, \citenamefont {Nishihara},
  \citenamefont {Inoue}, \citenamefont {Ovchinnikov},\ and\ \citenamefont
  {Kishine}}]{togawa2012chiral}%
  \BibitemOpen
  \bibfield  {author} {\bibinfo {author} {\bibfnamefont {Y.}~\bibnamefont
  {Togawa}}, \bibinfo {author} {\bibfnamefont {T.}~\bibnamefont {Koyama}},
  \bibinfo {author} {\bibfnamefont {K.}~\bibnamefont {Takayanagi}}, \bibinfo
  {author} {\bibfnamefont {S.}~\bibnamefont {Mori}}, \bibinfo {author}
  {\bibfnamefont {Y.}~\bibnamefont {Kousaka}}, \bibinfo {author} {\bibfnamefont
  {J.}~\bibnamefont {Akimitsu}}, \bibinfo {author} {\bibfnamefont
  {S.}~\bibnamefont {Nishihara}}, \bibinfo {author} {\bibfnamefont
  {K.}~\bibnamefont {Inoue}}, \bibinfo {author} {\bibfnamefont
  {A.}~\bibnamefont {Ovchinnikov}},\ and\ \bibinfo {author} {\bibfnamefont
  {J.-i.}\ \bibnamefont {Kishine}},\ }\bibfield  {title} {\bibinfo {title}
  {Chiral magnetic soliton lattice on a chiral helimagnet},\ }\href@noop {}
  {\bibfield  {journal} {\bibinfo  {journal} {Phys. Rev. Lett.}\ }\textbf
  {\bibinfo {volume} {108}},\ \bibinfo {pages} {107202} (\bibinfo {year}
  {2012})}\BibitemShut {NoStop}%
\bibitem [{\citenamefont {Kousaka}\ \emph {et~al.}(2016)\citenamefont
  {Kousaka}, \citenamefont {Ogura}, \citenamefont {Zhang}, \citenamefont
  {Miao}, \citenamefont {Lee}, \citenamefont {Torii}, \citenamefont {Kamiyama},
  \citenamefont {Campo}, \citenamefont {Inoue},\ and\ \citenamefont
  {Akimitsu}}]{Kousaka16}%
  \BibitemOpen
  \bibfield  {author} {\bibinfo {author} {\bibfnamefont {Y.}~\bibnamefont
  {Kousaka}}, \bibinfo {author} {\bibfnamefont {T.}~\bibnamefont {Ogura}},
  \bibinfo {author} {\bibfnamefont {J.}~\bibnamefont {Zhang}}, \bibinfo
  {author} {\bibfnamefont {P.}~\bibnamefont {Miao}}, \bibinfo {author}
  {\bibfnamefont {S.}~\bibnamefont {Lee}}, \bibinfo {author} {\bibfnamefont
  {S.}~\bibnamefont {Torii}}, \bibinfo {author} {\bibfnamefont
  {T.}~\bibnamefont {Kamiyama}}, \bibinfo {author} {\bibfnamefont
  {J.}~\bibnamefont {Campo}}, \bibinfo {author} {\bibfnamefont
  {K.}~\bibnamefont {Inoue}},\ and\ \bibinfo {author} {\bibfnamefont
  {J.}~\bibnamefont {Akimitsu}},\ }\bibfield  {title} {\bibinfo {title} {Long
  periodic helimagnetic ordering in {CrM$_3$S$_6$} {(M = Nb and Ta)}},\
  }\href@noop {} {\bibfield  {journal} {\bibinfo  {journal} {J. Phys.: Conf.
  Ser.}\ }\textbf {\bibinfo {volume} {746}},\ \bibinfo {pages} {012061}
  (\bibinfo {year} {2016})}\BibitemShut {NoStop}%
\bibitem [{\citenamefont {Zheludev}\ \emph {et~al.}(1997)\citenamefont
  {Zheludev}, \citenamefont {Maslov}, \citenamefont {Shirane}, \citenamefont
  {Sasago}, \citenamefont {Koide},\ and\ \citenamefont
  {Uchinokura}}]{Zheludev97}%
  \BibitemOpen
  \bibfield  {author} {\bibinfo {author} {\bibfnamefont {A.}~\bibnamefont
  {Zheludev}}, \bibinfo {author} {\bibfnamefont {S.}~\bibnamefont {Maslov}},
  \bibinfo {author} {\bibfnamefont {G.}~\bibnamefont {Shirane}}, \bibinfo
  {author} {\bibfnamefont {Y.}~\bibnamefont {Sasago}}, \bibinfo {author}
  {\bibfnamefont {N.}~\bibnamefont {Koide}},\ and\ \bibinfo {author}
  {\bibfnamefont {K.}~\bibnamefont {Uchinokura}},\ }\bibfield  {title}
  {\bibinfo {title} {Field-induced commensurate-incommensurate phase transition
  in a {Dzyaloshinskii-Moriya} spiral antiferromagnet},\ }\href@noop {}
  {\bibfield  {journal} {\bibinfo  {journal} {Phys. Rev. Lett.}\ }\textbf
  {\bibinfo {volume} {78}},\ \bibinfo {pages} {4857} (\bibinfo {year}
  {1997})}\BibitemShut {NoStop}%
\bibitem [{\citenamefont {Ghimire}\ \emph {et~al.}(2013)\citenamefont
  {Ghimire}, \citenamefont {McGuire}, \citenamefont {Parker}, \citenamefont
  {Sipos}, \citenamefont {Tang}, \citenamefont {Yan}, \citenamefont {Sales},\
  and\ \citenamefont {Mandrus}}]{Ghimire13}%
  \BibitemOpen
  \bibfield  {author} {\bibinfo {author} {\bibfnamefont {N.}~\bibnamefont
  {Ghimire}}, \bibinfo {author} {\bibfnamefont {M.}~\bibnamefont {McGuire}},
  \bibinfo {author} {\bibfnamefont {D.}~\bibnamefont {Parker}}, \bibinfo
  {author} {\bibfnamefont {B.}~\bibnamefont {Sipos}}, \bibinfo {author}
  {\bibfnamefont {S.}~\bibnamefont {Tang}}, \bibinfo {author} {\bibfnamefont
  {J.-Q.}\ \bibnamefont {Yan}}, \bibinfo {author} {\bibfnamefont
  {B.}~\bibnamefont {Sales}},\ and\ \bibinfo {author} {\bibfnamefont
  {D.}~\bibnamefont {Mandrus}},\ }\bibfield  {title} {\bibinfo {title}
  {Magnetic phase transition in single crystals of the chiral helimagnet
  {Cr$_{1/3}$NbS$_2$}},\ }\href@noop {} {\bibfield  {journal} {\bibinfo
  {journal} {Phys. Rev. B}\ }\textbf {\bibinfo {volume} {87}},\ \bibinfo
  {pages} {104403} (\bibinfo {year} {2013})}\BibitemShut {NoStop}%
\bibitem [{\citenamefont {Togawa}\ \emph {et~al.}(2013)\citenamefont {Togawa},
  \citenamefont {Kousaka}, \citenamefont {Nishihara}, \citenamefont {Inoue},
  \citenamefont {Akimitsu}, \citenamefont {Ovchinnikov},\ and\ \citenamefont
  {Kishine}}]{Togawa13}%
  \BibitemOpen
  \bibfield  {author} {\bibinfo {author} {\bibfnamefont {Y.}~\bibnamefont
  {Togawa}}, \bibinfo {author} {\bibfnamefont {Y.}~\bibnamefont {Kousaka}},
  \bibinfo {author} {\bibfnamefont {S.}~\bibnamefont {Nishihara}}, \bibinfo
  {author} {\bibfnamefont {K.}~\bibnamefont {Inoue}}, \bibinfo {author}
  {\bibfnamefont {J.}~\bibnamefont {Akimitsu}}, \bibinfo {author}
  {\bibfnamefont {A.}~\bibnamefont {Ovchinnikov}},\ and\ \bibinfo {author}
  {\bibfnamefont {J.}~\bibnamefont {Kishine}},\ }\bibfield  {title} {\bibinfo
  {title} {Interlayer magnetoresistance due to chiral soliton lattice formation
  in hexagonal chiral magnet {CrNb$_3$S$_6$}},\ }\href@noop {} {\bibfield
  {journal} {\bibinfo  {journal} {Phys. Rev. Lett.}\ }\textbf {\bibinfo
  {volume} {111}},\ \bibinfo {pages} {197204} (\bibinfo {year}
  {2013})}\BibitemShut {NoStop}%
\bibitem [{\citenamefont {Chapman}\ \emph {et~al.}(2014)\citenamefont
  {Chapman}, \citenamefont {Bornstein}, \citenamefont {Ghimire}, \citenamefont
  {Mandrus},\ and\ \citenamefont {Lee}}]{Chapman14}%
  \BibitemOpen
  \bibfield  {author} {\bibinfo {author} {\bibfnamefont {B.}~\bibnamefont
  {Chapman}}, \bibinfo {author} {\bibfnamefont {A.}~\bibnamefont {Bornstein}},
  \bibinfo {author} {\bibfnamefont {N.}~\bibnamefont {Ghimire}}, \bibinfo
  {author} {\bibfnamefont {D.}~\bibnamefont {Mandrus}},\ and\ \bibinfo {author}
  {\bibfnamefont {M.}~\bibnamefont {Lee}},\ }\bibfield  {title} {\bibinfo
  {title} {Spin structure of the anisotropic helimagnet {Cr$_{1/3}$NbS$_2$} in
  a magnetic field},\ }\href@noop {} {\bibfield  {journal} {\bibinfo  {journal}
  {Appl. Phys. Lett.}\ }\textbf {\bibinfo {volume} {105}},\ \bibinfo {pages}
  {072405} (\bibinfo {year} {2014})}\BibitemShut {NoStop}%
\bibitem [{\citenamefont {Kishine}\ and\ \citenamefont
  {Ovchinnikov}(2015)}]{Kishine15}%
  \BibitemOpen
  \bibfield  {author} {\bibinfo {author} {\bibfnamefont {J.}~\bibnamefont
  {Kishine}}\ and\ \bibinfo {author} {\bibfnamefont {A.}~\bibnamefont
  {Ovchinnikov}},\ }\bibfield  {title} {\bibinfo {title} {Theory of monoaxial
  chiral helimagnet},\ }\href@noop {} {\bibfield  {journal} {\bibinfo
  {journal} {Solid State Phys.}\ }\textbf {\bibinfo {volume} {66}},\ \bibinfo
  {pages} {1} (\bibinfo {year} {2015})}\BibitemShut {NoStop}%
\bibitem [{\citenamefont {Togawa}\ \emph {et~al.}(2016)\citenamefont {Togawa},
  \citenamefont {Kousaka}, \citenamefont {Inoue},\ and\ \citenamefont
  {Kishine}}]{Togawa16}%
  \BibitemOpen
  \bibfield  {author} {\bibinfo {author} {\bibfnamefont {Y.}~\bibnamefont
  {Togawa}}, \bibinfo {author} {\bibfnamefont {Y.}~\bibnamefont {Kousaka}},
  \bibinfo {author} {\bibfnamefont {K.}~\bibnamefont {Inoue}},\ and\ \bibinfo
  {author} {\bibfnamefont {J.}~\bibnamefont {Kishine}},\ }\bibfield  {title}
  {\bibinfo {title} {Symmetry, structure, and dynamics of monoaxial chiral
  magnets},\ }\href@noop {} {\bibfield  {journal} {\bibinfo  {journal} {J.
  Phys. Soc. Jpn.}\ }\textbf {\bibinfo {volume} {85}},\ \bibinfo {pages}
  {112001} (\bibinfo {year} {2016})}\BibitemShut {NoStop}%
\bibitem [{\citenamefont {Shinozaki}\ \emph {et~al.}(2016)\citenamefont
  {Shinozaki}, \citenamefont {Hoshino}, \citenamefont {Masaki}, \citenamefont
  {Kishine},\ and\ \citenamefont {Kato}}]{Shinozaki16}%
  \BibitemOpen
  \bibfield  {author} {\bibinfo {author} {\bibfnamefont {M.}~\bibnamefont
  {Shinozaki}}, \bibinfo {author} {\bibfnamefont {S.}~\bibnamefont {Hoshino}},
  \bibinfo {author} {\bibfnamefont {Y.}~\bibnamefont {Masaki}}, \bibinfo
  {author} {\bibfnamefont {J.}~\bibnamefont {Kishine}},\ and\ \bibinfo {author}
  {\bibfnamefont {Y.}~\bibnamefont {Kato}},\ }\bibfield  {title} {\bibinfo
  {title} {Finite-temperature properties of three-dimensional chiral
  helimagnets},\ }\href@noop {} {\bibfield  {journal} {\bibinfo  {journal} {J.
  Phys. Soc. Jpn.}\ }\textbf {\bibinfo {volume} {85}},\ \bibinfo {pages}
  {074710} (\bibinfo {year} {2016})}\BibitemShut {NoStop}%
\bibitem [{\citenamefont {Nishikawa}\ and\ \citenamefont
  {Hukushima}(2016)}]{Nishikawa16}%
  \BibitemOpen
  \bibfield  {author} {\bibinfo {author} {\bibfnamefont {Y.}~\bibnamefont
  {Nishikawa}}\ and\ \bibinfo {author} {\bibfnamefont {K.}~\bibnamefont
  {Hukushima}},\ }\bibfield  {title} {\bibinfo {title} {Phase transitions and
  ordering structures of a model of chiral helimagnet in three dimensions},\
  }\href@noop {} {\bibfield  {journal} {\bibinfo  {journal} {Phys. Rev. B}\
  }\textbf {\bibinfo {volume} {94}},\ \bibinfo {pages} {064428} (\bibinfo
  {year} {2016})}\BibitemShut {NoStop}%
\bibitem [{\citenamefont {Laliena}\ \emph
  {et~al.}(2016{\natexlab{a}})\citenamefont {Laliena}, \citenamefont {Campo},
  \citenamefont {Kishine}, \citenamefont {Ovchinnikov}, \citenamefont {Togawa},
  \citenamefont {Kousaka},\ and\ \citenamefont {Inoue}}]{Laliena16a}%
  \BibitemOpen
  \bibfield  {author} {\bibinfo {author} {\bibfnamefont {V.}~\bibnamefont
  {Laliena}}, \bibinfo {author} {\bibfnamefont {J.}~\bibnamefont {Campo}},
  \bibinfo {author} {\bibfnamefont {J.}~\bibnamefont {Kishine}}, \bibinfo
  {author} {\bibfnamefont {A.}~\bibnamefont {Ovchinnikov}}, \bibinfo {author}
  {\bibfnamefont {Y.}~\bibnamefont {Togawa}}, \bibinfo {author} {\bibfnamefont
  {Y.}~\bibnamefont {Kousaka}},\ and\ \bibinfo {author} {\bibfnamefont
  {K.}~\bibnamefont {Inoue}},\ }\bibfield  {title} {\bibinfo {title}
  {Incommensurate-commensurate transitions in the mono-axial chiral helimagnet
  driven by the magnetic field},\ }\href@noop {} {\bibfield  {journal}
  {\bibinfo  {journal} {Phys. Rev. B}\ }\textbf {\bibinfo {volume} {93}},\
  \bibinfo {pages} {134424} (\bibinfo {year} {2016}{\natexlab{a}})}\BibitemShut
  {NoStop}%
\bibitem [{\citenamefont {Laliena}\ \emph
  {et~al.}(2016{\natexlab{b}})\citenamefont {Laliena}, \citenamefont {Campo},\
  and\ \citenamefont {Kousaka}}]{Laliena16b}%
  \BibitemOpen
  \bibfield  {author} {\bibinfo {author} {\bibfnamefont {V.}~\bibnamefont
  {Laliena}}, \bibinfo {author} {\bibfnamefont {J.}~\bibnamefont {Campo}},\
  and\ \bibinfo {author} {\bibfnamefont {Y.}~\bibnamefont {Kousaka}},\
  }\bibfield  {title} {\bibinfo {title} {Understanding the {$H$-$T$} phase
  diagram of the monoaxial helimagnet},\ }\href@noop {} {\bibfield  {journal}
  {\bibinfo  {journal} {Phys. Rev. B}\ }\textbf {\bibinfo {volume} {94}},\
  \bibinfo {pages} {094439} (\bibinfo {year} {2016}{\natexlab{b}})}\BibitemShut
  {NoStop}%
\bibitem [{\citenamefont {Tsuruta}\ \emph {et~al.}(2016)\citenamefont
  {Tsuruta}, \citenamefont {Mito}, \citenamefont {Deguchi}, \citenamefont
  {Kishine}, \citenamefont {Kousaka}, \citenamefont {Akimitsu},\ and\
  \citenamefont {Inoue}}]{Tsuruta16}%
  \BibitemOpen
  \bibfield  {author} {\bibinfo {author} {\bibfnamefont {K.}~\bibnamefont
  {Tsuruta}}, \bibinfo {author} {\bibfnamefont {M.}~\bibnamefont {Mito}},
  \bibinfo {author} {\bibfnamefont {H.}~\bibnamefont {Deguchi}}, \bibinfo
  {author} {\bibfnamefont {J.}~\bibnamefont {Kishine}}, \bibinfo {author}
  {\bibfnamefont {Y.}~\bibnamefont {Kousaka}}, \bibinfo {author} {\bibfnamefont
  {J.}~\bibnamefont {Akimitsu}},\ and\ \bibinfo {author} {\bibfnamefont
  {K.}~\bibnamefont {Inoue}},\ }\bibfield  {title} {\bibinfo {title} {Phase
  diagram of the chiral magnet {Cr$_{1/3}$NbS$_2$} in a magnetic field},\
  }\href@noop {} {\bibfield  {journal} {\bibinfo  {journal} {Phys. Rev. B}\
  }\textbf {\bibinfo {volume} {93}},\ \bibinfo {pages} {104402} (\bibinfo
  {year} {2016})}\BibitemShut {NoStop}%
\bibitem [{\citenamefont {Han}\ \emph {et~al.}(2017)\citenamefont {Han},
  \citenamefont {Zhang}, \citenamefont {Sapkota}, \citenamefont {Hao},
  \citenamefont {Ling}, \citenamefont {Du}, \citenamefont {Pi}, \citenamefont
  {Zhang}, \citenamefont {Mandrus},\ and\ \citenamefont
  {Zhang}}]{han2017tricritical}%
  \BibitemOpen
  \bibfield  {author} {\bibinfo {author} {\bibfnamefont {H.}~\bibnamefont
  {Han}}, \bibinfo {author} {\bibfnamefont {L.}~\bibnamefont {Zhang}}, \bibinfo
  {author} {\bibfnamefont {D.}~\bibnamefont {Sapkota}}, \bibinfo {author}
  {\bibfnamefont {N.}~\bibnamefont {Hao}}, \bibinfo {author} {\bibfnamefont
  {L.}~\bibnamefont {Ling}}, \bibinfo {author} {\bibfnamefont {H.}~\bibnamefont
  {Du}}, \bibinfo {author} {\bibfnamefont {L.}~\bibnamefont {Pi}}, \bibinfo
  {author} {\bibfnamefont {C.}~\bibnamefont {Zhang}}, \bibinfo {author}
  {\bibfnamefont {D.~G.}\ \bibnamefont {Mandrus}},\ and\ \bibinfo {author}
  {\bibfnamefont {Y.}~\bibnamefont {Zhang}},\ }\bibfield  {title} {\bibinfo
  {title} {Tricritical point and phase diagram based on critical scaling in the
  monoaxial chiral helimagnet {Cr$_{1/3}$NbS$_2$}},\ }\href@noop {} {\bibfield
  {journal} {\bibinfo  {journal} {Phys. Rev. B}\ }\textbf {\bibinfo {volume}
  {96}},\ \bibinfo {pages} {094439} (\bibinfo {year} {2017})}\BibitemShut
  {NoStop}%
\bibitem [{\citenamefont {Yonemura}\ \emph {et~al.}(2017)\citenamefont
  {Yonemura}, \citenamefont {Shimamoto}, \citenamefont {Kida}, \citenamefont
  {Yoshizawa}, \citenamefont {Kousaka}, \citenamefont {Nishihara},
  \citenamefont {Goncalves}, \citenamefont {Akimitsu}, \citenamefont {Inoue},
  \citenamefont {Hagiwara},\ and\ \citenamefont
  {Togawa}}]{yonemura2017magnetic}%
  \BibitemOpen
  \bibfield  {author} {\bibinfo {author} {\bibfnamefont {J.-i.}\ \bibnamefont
  {Yonemura}}, \bibinfo {author} {\bibfnamefont {Y.}~\bibnamefont {Shimamoto}},
  \bibinfo {author} {\bibfnamefont {T.}~\bibnamefont {Kida}}, \bibinfo {author}
  {\bibfnamefont {D.}~\bibnamefont {Yoshizawa}}, \bibinfo {author}
  {\bibfnamefont {Y.}~\bibnamefont {Kousaka}}, \bibinfo {author} {\bibfnamefont
  {S.}~\bibnamefont {Nishihara}}, \bibinfo {author} {\bibfnamefont {F.~J.~T.}\
  \bibnamefont {Goncalves}}, \bibinfo {author} {\bibfnamefont {J.}~\bibnamefont
  {Akimitsu}}, \bibinfo {author} {\bibfnamefont {K.}~\bibnamefont {Inoue}},
  \bibinfo {author} {\bibfnamefont {M.}~\bibnamefont {Hagiwara}},\ and\
  \bibinfo {author} {\bibfnamefont {Y.}~\bibnamefont {Togawa}},\ }\bibfield
  {title} {\bibinfo {title} {Magnetic solitons and magnetic phase diagram of
  the hexagonal chiral crystal {CrNb$_3$S$_6$} in oblique magnetic fields},\
  }\href@noop {} {\bibfield  {journal} {\bibinfo  {journal} {Phys. Rev. B}\
  }\textbf {\bibinfo {volume} {96}},\ \bibinfo {pages} {184423} (\bibinfo
  {year} {2017})}\BibitemShut {NoStop}%
\bibitem [{\citenamefont {Clements}\ \emph {et~al.}(2017)\citenamefont
  {Clements}, \citenamefont {Das}, \citenamefont {Li}, \citenamefont
  {Lampen-Kelley}, \citenamefont {Phan}, \citenamefont {Keppens}, \citenamefont
  {Mandrus},\ and\ \citenamefont {Srikanth}}]{clements2017critical}%
  \BibitemOpen
  \bibfield  {author} {\bibinfo {author} {\bibfnamefont {E.~M.}\ \bibnamefont
  {Clements}}, \bibinfo {author} {\bibfnamefont {R.}~\bibnamefont {Das}},
  \bibinfo {author} {\bibfnamefont {L.}~\bibnamefont {Li}}, \bibinfo {author}
  {\bibfnamefont {P.~J.}\ \bibnamefont {Lampen-Kelley}}, \bibinfo {author}
  {\bibfnamefont {M.-H.}\ \bibnamefont {Phan}}, \bibinfo {author}
  {\bibfnamefont {V.}~\bibnamefont {Keppens}}, \bibinfo {author} {\bibfnamefont
  {D.}~\bibnamefont {Mandrus}},\ and\ \bibinfo {author} {\bibfnamefont
  {H.}~\bibnamefont {Srikanth}},\ }\bibfield  {title} {\bibinfo {title}
  {Critical behavior and macroscopic phase diagram of the monoaxial chiral
  helimagnet {Cr$_{1/3}$NbS$_2$}},\ }\href@noop {} {\bibfield  {journal}
  {\bibinfo  {journal} {Sci. Rep.}\ }\textbf {\bibinfo {volume} {7}},\ \bibinfo
  {pages} {1} (\bibinfo {year} {2017})}\BibitemShut {NoStop}%
\bibitem [{\citenamefont {Laliena}\ \emph {et~al.}(2017)\citenamefont
  {Laliena}, \citenamefont {Campo},\ and\ \citenamefont
  {Kousaka}}]{Laliena17a}%
  \BibitemOpen
  \bibfield  {author} {\bibinfo {author} {\bibfnamefont {V.}~\bibnamefont
  {Laliena}}, \bibinfo {author} {\bibfnamefont {J.}~\bibnamefont {Campo}},\
  and\ \bibinfo {author} {\bibfnamefont {Y.}~\bibnamefont {Kousaka}},\
  }\bibfield  {title} {\bibinfo {title} {Nucleation, instability, and
  discontinuous phase transitions in the phase diagram of the monoaxial
  helimagnet with oblique fields},\ }\href@noop {} {\bibfield  {journal}
  {\bibinfo  {journal} {Phys. Rev. B}\ }\textbf {\bibinfo {volume} {95}},\
  \bibinfo {pages} {224410} (\bibinfo {year} {2017})}\BibitemShut {NoStop}%
\bibitem [{\citenamefont {Masaki}\ \emph {et~al.}(2018)\citenamefont {Masaki},
  \citenamefont {Aoki}, \citenamefont {Togawa},\ and\ \citenamefont
  {Kato}}]{Masaki18}%
  \BibitemOpen
  \bibfield  {author} {\bibinfo {author} {\bibfnamefont {Y.}~\bibnamefont
  {Masaki}}, \bibinfo {author} {\bibfnamefont {R.}~\bibnamefont {Aoki}},
  \bibinfo {author} {\bibfnamefont {Y.}~\bibnamefont {Togawa}},\ and\ \bibinfo
  {author} {\bibfnamefont {Y.}~\bibnamefont {Kato}},\ }\bibfield  {title}
  {\bibinfo {title} {Chiral solitons in monoaxial chiral magnets in tilted
  magnetic field},\ }\href@noop {} {\bibfield  {journal} {\bibinfo  {journal}
  {Phys. Rev. B}\ }\textbf {\bibinfo {volume} {98}},\ \bibinfo {pages}
  {100402(R)} (\bibinfo {year} {2018})}\BibitemShut {NoStop}%
\bibitem [{\citenamefont {Laliena}\ \emph {et~al.}(2018)\citenamefont
  {Laliena}, \citenamefont {Kato}, \citenamefont {Albalate},\ and\
  \citenamefont {Campo}}]{Laliena18a}%
  \BibitemOpen
  \bibfield  {author} {\bibinfo {author} {\bibfnamefont {V.}~\bibnamefont
  {Laliena}}, \bibinfo {author} {\bibfnamefont {Y.}~\bibnamefont {Kato}},
  \bibinfo {author} {\bibfnamefont {G.}~\bibnamefont {Albalate}},\ and\
  \bibinfo {author} {\bibfnamefont {J.}~\bibnamefont {Campo}},\ }\bibfield
  {title} {\bibinfo {title} {Thermal fluctuations in the conical state of
  monoaxial helimagnets},\ }\href@noop {} {\bibfield  {journal} {\bibinfo
  {journal} {Phys. Rev. B}\ }\textbf {\bibinfo {volume} {98}},\ \bibinfo
  {pages} {144445} (\bibinfo {year} {2018})}\BibitemShut {NoStop}%
\bibitem [{\citenamefont {Aoki}\ \emph {et~al.}(2019)\citenamefont {Aoki},
  \citenamefont {Kousaka},\ and\ \citenamefont {Togawa}}]{aoki2019anomalous}%
  \BibitemOpen
  \bibfield  {author} {\bibinfo {author} {\bibfnamefont {R.}~\bibnamefont
  {Aoki}}, \bibinfo {author} {\bibfnamefont {Y.}~\bibnamefont {Kousaka}},\ and\
  \bibinfo {author} {\bibfnamefont {Y.}~\bibnamefont {Togawa}},\ }\bibfield
  {title} {\bibinfo {title} {Anomalous nonreciprocal electrical transport on
  chiral magnetic order},\ }\href@noop {} {\bibfield  {journal} {\bibinfo
  {journal} {Phys. Rev. Lett.}\ }\textbf {\bibinfo {volume} {122}},\ \bibinfo
  {pages} {057206} (\bibinfo {year} {2019})}\BibitemShut {NoStop}%
\bibitem [{\citenamefont {Honda}\ \emph {et~al.}(2020)\citenamefont {Honda},
  \citenamefont {Yamasaki}, \citenamefont {Nakao}, \citenamefont {Murakami},
  \citenamefont {Ogura}, \citenamefont {Kousaka},\ and\ \citenamefont
  {Akimitsu}}]{honda2020topological}%
  \BibitemOpen
  \bibfield  {author} {\bibinfo {author} {\bibfnamefont {T.}~\bibnamefont
  {Honda}}, \bibinfo {author} {\bibfnamefont {Y.}~\bibnamefont {Yamasaki}},
  \bibinfo {author} {\bibfnamefont {H.}~\bibnamefont {Nakao}}, \bibinfo
  {author} {\bibfnamefont {Y.}~\bibnamefont {Murakami}}, \bibinfo {author}
  {\bibfnamefont {T.}~\bibnamefont {Ogura}}, \bibinfo {author} {\bibfnamefont
  {Y.}~\bibnamefont {Kousaka}},\ and\ \bibinfo {author} {\bibfnamefont
  {J.}~\bibnamefont {Akimitsu}},\ }\bibfield  {title} {\bibinfo {title}
  {Topological metastability supported by thermal fluctuation upon formation of
  chiral soliton lattice in {CrNb$_3$S$_6$}},\ }\href@noop {} {\bibfield
  {journal} {\bibinfo  {journal} {Sci. Rep.}\ }\textbf {\bibinfo {volume}
  {10}},\ \bibinfo {pages} {1} (\bibinfo {year} {2020})}\BibitemShut {NoStop}%
\bibitem [{\citenamefont {Masaki}(2020)}]{Masaki20}%
  \BibitemOpen
  \bibfield  {author} {\bibinfo {author} {\bibfnamefont {Y.}~\bibnamefont
  {Masaki}},\ }\bibfield  {title} {\bibinfo {title} {Instabilities in monoaxial
  chiral magnets under a tilted magnetic field},\ }\href
  {https://doi.org/10.1103/PhysRevB.101.214424} {\bibfield  {journal} {\bibinfo
   {journal} {Phys. Rev. B}\ }\textbf {\bibinfo {volume} {101}},\ \bibinfo
  {pages} {214424} (\bibinfo {year} {2020})}\BibitemShut {NoStop}%
\bibitem [{\citenamefont {Karna}\ \emph {et~al.}(2021)\citenamefont {Karna},
  \citenamefont {Marshall}, \citenamefont {Xie}, \citenamefont
  {{DeBeer-Schmitt}}, \citenamefont {Young}, \citenamefont {Vekhter},
  \citenamefont {Shelton}, \citenamefont {Kov{\'a}cs}, \citenamefont
  {Charilaou},\ and\ \citenamefont {DiTusa}}]{Karna2021}%
  \BibitemOpen
  \bibfield  {author} {\bibinfo {author} {\bibfnamefont {S.~K.}\ \bibnamefont
  {Karna}}, \bibinfo {author} {\bibfnamefont {M.}~\bibnamefont {Marshall}},
  \bibinfo {author} {\bibfnamefont {W.}~\bibnamefont {Xie}}, \bibinfo {author}
  {\bibfnamefont {L.}~\bibnamefont {{DeBeer-Schmitt}}}, \bibinfo {author}
  {\bibfnamefont {D.~P.}\ \bibnamefont {Young}}, \bibinfo {author}
  {\bibfnamefont {I.}~\bibnamefont {Vekhter}}, \bibinfo {author} {\bibfnamefont
  {W.~A.}\ \bibnamefont {Shelton}}, \bibinfo {author} {\bibfnamefont
  {A.}~\bibnamefont {Kov{\'a}cs}}, \bibinfo {author} {\bibfnamefont
  {M.}~\bibnamefont {Charilaou}},\ and\ \bibinfo {author} {\bibfnamefont
  {J.~F.}\ \bibnamefont {DiTusa}},\ }\bibfield  {title} {\bibinfo {title}
  {Annihilation and control of chiral domain walls with magnetic fields},\
  }\href@noop {} {\bibfield  {journal} {\bibinfo  {journal} {Nano Lett.}\
  }\textbf {\bibinfo {volume} {21}},\ \bibinfo {pages} {1205} (\bibinfo {year}
  {2021})}\BibitemShut {NoStop}%
\bibitem [{\citenamefont {Mankovsky}\ \emph {et~al.}(2016)\citenamefont
  {Mankovsky}, \citenamefont {Polesya}, \citenamefont {Ebert},\ and\
  \citenamefont {Bensch}}]{mankovsky2016electronic}%
  \BibitemOpen
  \bibfield  {author} {\bibinfo {author} {\bibfnamefont {S.}~\bibnamefont
  {Mankovsky}}, \bibinfo {author} {\bibfnamefont {S.}~\bibnamefont {Polesya}},
  \bibinfo {author} {\bibfnamefont {H.}~\bibnamefont {Ebert}},\ and\ \bibinfo
  {author} {\bibfnamefont {W.}~\bibnamefont {Bensch}},\ }\bibfield  {title}
  {\bibinfo {title} {Electronic and magnetic properties of {$2 H$-- NbS$_2$}
  intercalated by 3 d transition metals},\ }\href@noop {} {\bibfield  {journal}
  {\bibinfo  {journal} {Physical Review B}\ }\textbf {\bibinfo {volume} {94}},\
  \bibinfo {pages} {184430} (\bibinfo {year} {2016})}\BibitemShut {NoStop}%
\bibitem [{\citenamefont {Kousaka}\ \emph {et~al.}(2009)\citenamefont
  {Kousaka}, \citenamefont {Nakao}, \citenamefont {Kishine}, \citenamefont
  {Akita}, \citenamefont {Inoue},\ and\ \citenamefont
  {Akimitsu}}]{kousaka2009chiral}%
  \BibitemOpen
  \bibfield  {author} {\bibinfo {author} {\bibfnamefont {Y.}~\bibnamefont
  {Kousaka}}, \bibinfo {author} {\bibfnamefont {Y.}~\bibnamefont {Nakao}},
  \bibinfo {author} {\bibfnamefont {J.-i.}\ \bibnamefont {Kishine}}, \bibinfo
  {author} {\bibfnamefont {M.}~\bibnamefont {Akita}}, \bibinfo {author}
  {\bibfnamefont {K.}~\bibnamefont {Inoue}},\ and\ \bibinfo {author}
  {\bibfnamefont {J.}~\bibnamefont {Akimitsu}},\ }\bibfield  {title} {\bibinfo
  {title} {Chiral helimagnetism in {T$_{1/3}$NbS$_2$ (T= Cr and Mn)}},\
  }\href@noop {} {\bibfield  {journal} {\bibinfo  {journal} {Nucl. Instrum.
  Methods Phys. Res. Sect. A}\ }\textbf {\bibinfo {volume} {600}},\ \bibinfo
  {pages} {250} (\bibinfo {year} {2009})}\BibitemShut {NoStop}%
\bibitem [{\citenamefont {Karna}\ \emph {et~al.}(2019)\citenamefont {Karna},
  \citenamefont {Womack}, \citenamefont {Chapai}, \citenamefont {Young},
  \citenamefont {Marshall}, \citenamefont {Xie}, \citenamefont {Graf},
  \citenamefont {Wu}, \citenamefont {Cao}, \citenamefont {DeBeer-Schmitt},
  \citenamefont {Adams}, \citenamefont {Jin},\ and\ \citenamefont
  {{DiTusa}}}]{karna2019consequences}%
  \BibitemOpen
  \bibfield  {author} {\bibinfo {author} {\bibfnamefont {S.~K.}\ \bibnamefont
  {Karna}}, \bibinfo {author} {\bibfnamefont {F.~N.}\ \bibnamefont {Womack}},
  \bibinfo {author} {\bibfnamefont {R.}~\bibnamefont {Chapai}}, \bibinfo
  {author} {\bibfnamefont {D.~P.}\ \bibnamefont {Young}}, \bibinfo {author}
  {\bibfnamefont {M.}~\bibnamefont {Marshall}}, \bibinfo {author}
  {\bibfnamefont {W.}~\bibnamefont {Xie}}, \bibinfo {author} {\bibfnamefont
  {D.}~\bibnamefont {Graf}}, \bibinfo {author} {\bibfnamefont {Y.}~\bibnamefont
  {Wu}}, \bibinfo {author} {\bibfnamefont {H.}~\bibnamefont {Cao}}, \bibinfo
  {author} {\bibfnamefont {L.}~\bibnamefont {DeBeer-Schmitt}}, \bibinfo
  {author} {\bibfnamefont {P.~W.}\ \bibnamefont {Adams}}, \bibinfo {author}
  {\bibfnamefont {R.}~\bibnamefont {Jin}},\ and\ \bibinfo {author}
  {\bibfnamefont {J.~F.}\ \bibnamefont {{DiTusa}}},\ }\bibfield  {title}
  {\bibinfo {title} {Consequences of magnetic ordering in chiral
  {Mn$_{1/3}$NbS$_2$}},\ }\href@noop {} {\bibfield  {journal} {\bibinfo
  {journal} {Phys. Rev. B}\ }\textbf {\bibinfo {volume} {100}},\ \bibinfo
  {pages} {184413} (\bibinfo {year} {2019})}\BibitemShut {NoStop}%
\bibitem [{\citenamefont {Dai}\ \emph {et~al.}(2019)\citenamefont {Dai},
  \citenamefont {Liu}, \citenamefont {Wang}, \citenamefont {Fan}, \citenamefont
  {Pi}, \citenamefont {Zhang},\ and\ \citenamefont {Zhang}}]{dai2019critical}%
  \BibitemOpen
  \bibfield  {author} {\bibinfo {author} {\bibfnamefont {Y.}~\bibnamefont
  {Dai}}, \bibinfo {author} {\bibfnamefont {W.}~\bibnamefont {Liu}}, \bibinfo
  {author} {\bibfnamefont {Y.}~\bibnamefont {Wang}}, \bibinfo {author}
  {\bibfnamefont {J.}~\bibnamefont {Fan}}, \bibinfo {author} {\bibfnamefont
  {L.}~\bibnamefont {Pi}}, \bibinfo {author} {\bibfnamefont {L.}~\bibnamefont
  {Zhang}},\ and\ \bibinfo {author} {\bibfnamefont {Y.}~\bibnamefont {Zhang}},\
  }\bibfield  {title} {\bibinfo {title} {Critical phenomenon and phase diagram
  of mn-intercalated layered {MnNb$_3$S$_6$}},\ }\href@noop {} {\bibfield
  {journal} {\bibinfo  {journal} {J. Phys.: Cond. Matter}\ }\textbf {\bibinfo
  {volume} {31}},\ \bibinfo {pages} {195803} (\bibinfo {year}
  {2019})}\BibitemShut {NoStop}%
\bibitem [{\citenamefont {Ohkuma}\ \emph {et~al.}(2022)\citenamefont {Ohkuma},
  \citenamefont {Mito}, \citenamefont {Deguchi}, \citenamefont {Kousaka},
  \citenamefont {Ohe}, \citenamefont {Akimitsu}, \citenamefont {Kishine},\ and\
  \citenamefont {Inoue}}]{ohkuma2022nonequilibrium}%
  \BibitemOpen
  \bibfield  {author} {\bibinfo {author} {\bibfnamefont {M.}~\bibnamefont
  {Ohkuma}}, \bibinfo {author} {\bibfnamefont {M.}~\bibnamefont {Mito}},
  \bibinfo {author} {\bibfnamefont {H.}~\bibnamefont {Deguchi}}, \bibinfo
  {author} {\bibfnamefont {Y.}~\bibnamefont {Kousaka}}, \bibinfo {author}
  {\bibfnamefont {J.}~\bibnamefont {Ohe}}, \bibinfo {author} {\bibfnamefont
  {J.}~\bibnamefont {Akimitsu}}, \bibinfo {author} {\bibfnamefont
  {J.}~\bibnamefont {Kishine}},\ and\ \bibinfo {author} {\bibfnamefont
  {K.}~\bibnamefont {Inoue}},\ }\bibfield  {title} {\bibinfo {title}
  {Nonequilibrium chiral soliton lattice in the monoaxial chiral magnet
  {MnNb$_3$S$_6$}},\ }\href@noop {} {\bibfield  {journal} {\bibinfo  {journal}
  {Phys. Rev. B}\ }\textbf {\bibinfo {volume} {106}},\ \bibinfo {pages}
  {104410} (\bibinfo {year} {2022})}\BibitemShut {NoStop}%
\bibitem [{\citenamefont {Li}\ \emph {et~al.}(2023)\citenamefont {Li},
  \citenamefont {Li}, \citenamefont {Zhou}, \citenamefont {Gu}, \citenamefont
  {Fu}, \citenamefont {Chen}, \citenamefont {Zhang},\ and\ \citenamefont
  {Liu}}]{li2023temperature}%
  \BibitemOpen
  \bibfield  {author} {\bibinfo {author} {\bibfnamefont {L.}~\bibnamefont
  {Li}}, \bibinfo {author} {\bibfnamefont {H.}~\bibnamefont {Li}}, \bibinfo
  {author} {\bibfnamefont {K.}~\bibnamefont {Zhou}}, \bibinfo {author}
  {\bibfnamefont {Y.}~\bibnamefont {Gu}}, \bibinfo {author} {\bibfnamefont
  {Q.}~\bibnamefont {Fu}}, \bibinfo {author} {\bibfnamefont {L.}~\bibnamefont
  {Chen}}, \bibinfo {author} {\bibfnamefont {L.}~\bibnamefont {Zhang}},\ and\
  \bibinfo {author} {\bibfnamefont {R.}~\bibnamefont {Liu}},\ }\bibfield
  {title} {\bibinfo {title} {Temperature-and field angular-dependent helical
  spin period characterized by magnetic dynamics in a chiral helimagnet
  {MnNb$_3$S$_6$}},\ }\href@noop {} {\bibfield  {journal} {\bibinfo  {journal}
  {Sci. China Phys. Mech. \& Ast.}\ }\textbf {\bibinfo {volume} {66}},\
  \bibinfo {pages} {217511} (\bibinfo {year} {2023})}\BibitemShut {NoStop}%
\bibitem [{\citenamefont {Hall}\ \emph {et~al.}(2022)\citenamefont {Hall},
  \citenamefont {Loudon}, \citenamefont {Midgley}, \citenamefont
  {Twitchett-Harrison}, \citenamefont {Holt}, \citenamefont {Mayoh},
  \citenamefont {Tidey}, \citenamefont {Han}, \citenamefont {Lees},\ and\
  \citenamefont {Balakrishnan}}]{hall2022comparative}%
  \BibitemOpen
  \bibfield  {author} {\bibinfo {author} {\bibfnamefont {A.}~\bibnamefont
  {Hall}}, \bibinfo {author} {\bibfnamefont {J.}~\bibnamefont {Loudon}},
  \bibinfo {author} {\bibfnamefont {P.}~\bibnamefont {Midgley}}, \bibinfo
  {author} {\bibfnamefont {A.}~\bibnamefont {Twitchett-Harrison}}, \bibinfo
  {author} {\bibfnamefont {S.}~\bibnamefont {Holt}}, \bibinfo {author}
  {\bibfnamefont {D.}~\bibnamefont {Mayoh}}, \bibinfo {author} {\bibfnamefont
  {J.}~\bibnamefont {Tidey}}, \bibinfo {author} {\bibfnamefont
  {Y.}~\bibnamefont {Han}}, \bibinfo {author} {\bibfnamefont {M.}~\bibnamefont
  {Lees}},\ and\ \bibinfo {author} {\bibfnamefont {G.}~\bibnamefont
  {Balakrishnan}},\ }\bibfield  {title} {\bibinfo {title} {Comparative study of
  the structural and magnetic properties of {Mn$_{1/3}$NbS$_2$} and
  {Cr$_{1/3}$NbS$_2$}},\ }\href@noop {} {\bibfield  {journal} {\bibinfo
  {journal} {Phys. Rev. Mat.}\ }\textbf {\bibinfo {volume} {6}},\ \bibinfo
  {pages} {024407} (\bibinfo {year} {2022})}\BibitemShut {NoStop}%
\bibitem [{\citenamefont {Laliena}\ \emph {et~al.}(2020)\citenamefont
  {Laliena}, \citenamefont {Bustingorry},\ and\ \citenamefont
  {Campo}}]{victor2020dynamics}%
  \BibitemOpen
  \bibfield  {author} {\bibinfo {author} {\bibfnamefont {V.}~\bibnamefont
  {Laliena}}, \bibinfo {author} {\bibfnamefont {S.}~\bibnamefont
  {Bustingorry}},\ and\ \bibinfo {author} {\bibfnamefont {J.}~\bibnamefont
  {Campo}},\ }\bibfield  {title} {\bibinfo {title} {Dynamics of chiral solitons
  driven by polarized currents in monoaxial helimagnets},\ }\href@noop {}
  {\bibfield  {journal} {\bibinfo  {journal} {Sci. Rep.}\ }\textbf {\bibinfo
  {volume} {10}},\ \bibinfo {pages} {20430} (\bibinfo {year}
  {2020})}\BibitemShut {NoStop}%
\bibitem [{\citenamefont {Osorio}\ \emph {et~al.}(2021)\citenamefont {Osorio},
  \citenamefont {Laliena}, \citenamefont {Campo},\ and\ \citenamefont
  {Bustingorry}}]{Osorio2021}%
  \BibitemOpen
  \bibfield  {author} {\bibinfo {author} {\bibfnamefont {S.~A.}\ \bibnamefont
  {Osorio}}, \bibinfo {author} {\bibfnamefont {V.}~\bibnamefont {Laliena}},
  \bibinfo {author} {\bibfnamefont {J.}~\bibnamefont {Campo}},\ and\ \bibinfo
  {author} {\bibfnamefont {S.}~\bibnamefont {Bustingorry}},\ }\bibfield
  {title} {\bibinfo {title} {Creation of single chiral soliton states in
  monoaxial helimagnets},\ }\href@noop {} {\bibfield  {journal} {\bibinfo
  {journal} {Appl. Phys. Lett.}\ }\textbf {\bibinfo {volume} {119}},\ \bibinfo
  {pages} {222405} (\bibinfo {year} {2021})}\BibitemShut {NoStop}%
\bibitem [{\citenamefont {Vansteenkiste}\ \emph {et~al.}(2014)\citenamefont
  {Vansteenkiste}, \citenamefont {Leliaert}, \citenamefont {Dvornik},
  \citenamefont {Helsen}, \citenamefont {Garcia-Sanchez},\ and\ \citenamefont
  {Waeyenberge}}]{MuMax3}%
  \BibitemOpen
  \bibfield  {author} {\bibinfo {author} {\bibfnamefont {A.}~\bibnamefont
  {Vansteenkiste}}, \bibinfo {author} {\bibfnamefont {J.}~\bibnamefont
  {Leliaert}}, \bibinfo {author} {\bibfnamefont {M.}~\bibnamefont {Dvornik}},
  \bibinfo {author} {\bibfnamefont {M.}~\bibnamefont {Helsen}}, \bibinfo
  {author} {\bibfnamefont {F.}~\bibnamefont {Garcia-Sanchez}},\ and\ \bibinfo
  {author} {\bibfnamefont {B.~V.}\ \bibnamefont {Waeyenberge}},\ }\bibfield
  {title} {\bibinfo {title} {The design and verification of mumax3},\ }\href
  {https://doi.org/http://dx.doi.org/10.1063/1.4899186} {\bibfield  {journal}
  {\bibinfo  {journal} {AIP Adv.}\ }\textbf {\bibinfo {volume} {4}},\ \bibinfo
  {pages} {107133} (\bibinfo {year} {2014})}\BibitemShut {NoStop}%
\bibitem [{\citenamefont {Leliaert}\ \emph {et~al.}(2018)\citenamefont
  {Leliaert}, \citenamefont {Dvornik}, \citenamefont {Mulkers}, \citenamefont
  {{De Clercq}}, \citenamefont {Milo\v{s}evi\'{c}},\ and\ \citenamefont {{Van
  Waeyenberge}}}]{Leliaert18}%
  \BibitemOpen
  \bibfield  {author} {\bibinfo {author} {\bibfnamefont {J.}~\bibnamefont
  {Leliaert}}, \bibinfo {author} {\bibfnamefont {M.}~\bibnamefont {Dvornik}},
  \bibinfo {author} {\bibfnamefont {J.}~\bibnamefont {Mulkers}}, \bibinfo
  {author} {\bibfnamefont {J.}~\bibnamefont {{De Clercq}}}, \bibinfo {author}
  {\bibfnamefont {M.~V.}\ \bibnamefont {Milo\v{s}evi\'{c}}},\ and\ \bibinfo
  {author} {\bibfnamefont {B.}~\bibnamefont {{Van Waeyenberge}}},\ }\bibfield
  {title} {\bibinfo {title} {Fast micromagnetic simulations on {GPU—recent}
  advances made with {mumax$^3$}},\ }\href
  {https://doi.org/https://doi.org/10.1088/1361-6463/aaab1c} {\bibfield
  {journal} {\bibinfo  {journal} {J. Phys D: Appl. Phys.}\ }\textbf {\bibinfo
  {volume} {51}},\ \bibinfo {pages} {123002} (\bibinfo {year}
  {2018})}\BibitemShut {NoStop}%
\bibitem [{\citenamefont {Beg}\ \emph {et~al.}(2022)\citenamefont {Beg},
  \citenamefont {Lang},\ and\ \citenamefont {Fangohr}}]{beg2022}%
  \BibitemOpen
  \bibfield  {author} {\bibinfo {author} {\bibfnamefont {M.}~\bibnamefont
  {Beg}}, \bibinfo {author} {\bibfnamefont {M.}~\bibnamefont {Lang}},\ and\
  \bibinfo {author} {\bibfnamefont {H.}~\bibnamefont {Fangohr}},\ }\bibfield
  {title} {\bibinfo {title} {Ubermag: Towards more effective micromagnetic
  workflows},\ }\href {https://doi.org/10.1109/TMAG.2021.3078896} {\bibfield
  {journal} {\bibinfo  {journal} {IEEE Transactions on Magnetics}\ }\textbf
  {\bibinfo {volume} {58}},\ \bibinfo {pages} {1} (\bibinfo {year}
  {2022})}\BibitemShut {NoStop}%
\bibitem [{\citenamefont {Togawa}\ \emph {et~al.}(2015)\citenamefont {Togawa},
  \citenamefont {Koyama}, \citenamefont {Nishimori}, \citenamefont {Matsumoto},
  \citenamefont {McVitie}, \citenamefont {McGrouther}, \citenamefont {Stamps},
  \citenamefont {Kousaka}, \citenamefont {Akimitsu}, \citenamefont {Nishihara}
  \emph {et~al.}}]{togawa2015magnetic}%
  \BibitemOpen
  \bibfield  {author} {\bibinfo {author} {\bibfnamefont {Y.}~\bibnamefont
  {Togawa}}, \bibinfo {author} {\bibfnamefont {T.}~\bibnamefont {Koyama}},
  \bibinfo {author} {\bibfnamefont {Y.}~\bibnamefont {Nishimori}}, \bibinfo
  {author} {\bibfnamefont {Y.}~\bibnamefont {Matsumoto}}, \bibinfo {author}
  {\bibfnamefont {S.}~\bibnamefont {McVitie}}, \bibinfo {author} {\bibfnamefont
  {D.}~\bibnamefont {McGrouther}}, \bibinfo {author} {\bibfnamefont
  {R.}~\bibnamefont {Stamps}}, \bibinfo {author} {\bibfnamefont
  {Y.}~\bibnamefont {Kousaka}}, \bibinfo {author} {\bibfnamefont
  {J.}~\bibnamefont {Akimitsu}}, \bibinfo {author} {\bibfnamefont
  {S.}~\bibnamefont {Nishihara}}, \emph {et~al.},\ }\bibfield  {title}
  {\bibinfo {title} {Magnetic soliton confinement and discretization effects
  arising from macroscopic coherence in a chiral spin soliton lattice},\
  }\href@noop {} {\bibfield  {journal} {\bibinfo  {journal} {Phys. Rev. B}\
  }\textbf {\bibinfo {volume} {92}},\ \bibinfo {pages} {220412} (\bibinfo
  {year} {2015})}\BibitemShut {NoStop}%
\bibitem [{\citenamefont {Dai}\ \emph {et~al.}(2020)\citenamefont {Dai},
  \citenamefont {Liu}, \citenamefont {Wang}, \citenamefont {Zhao},
  \citenamefont {Meng}, \citenamefont {Tong}, \citenamefont {Pi}, \citenamefont
  {Zhang},\ and\ \citenamefont {Zhang}}]{dai2020microwave}%
  \BibitemOpen
  \bibfield  {author} {\bibinfo {author} {\bibfnamefont {Y.}~\bibnamefont
  {Dai}}, \bibinfo {author} {\bibfnamefont {W.}~\bibnamefont {Liu}}, \bibinfo
  {author} {\bibfnamefont {Y.}~\bibnamefont {Wang}}, \bibinfo {author}
  {\bibfnamefont {J.}~\bibnamefont {Zhao}}, \bibinfo {author} {\bibfnamefont
  {F.}~\bibnamefont {Meng}}, \bibinfo {author} {\bibfnamefont {W.}~\bibnamefont
  {Tong}}, \bibinfo {author} {\bibfnamefont {L.}~\bibnamefont {Pi}}, \bibinfo
  {author} {\bibfnamefont {L.}~\bibnamefont {Zhang}},\ and\ \bibinfo {author}
  {\bibfnamefont {Y.}~\bibnamefont {Zhang}},\ }\bibfield  {title} {\bibinfo
  {title} {Microwave response of the chiral helimagnetic {MnNb$_3$S$_6$}},\
  }\href@noop {} {\bibfield  {journal} {\bibinfo  {journal} {Appl. Phys.
  Lett.}\ }\textbf {\bibinfo {volume} {117}},\ \bibinfo {pages} {022410}
  (\bibinfo {year} {2020})}\BibitemShut {NoStop}%
\bibitem [{\citenamefont {Bostrem}\ \emph {et~al.}(2008)\citenamefont
  {Bostrem}, \citenamefont {\mbox{Jun-ichiro} Kishine},\ and\ \citenamefont
  {Ovchinnikov}}]{Bostrem08b}%
  \BibitemOpen
  \bibfield  {author} {\bibinfo {author} {\bibfnamefont {I.}~\bibnamefont
  {Bostrem}}, \bibinfo {author} {\bibnamefont {\mbox{Jun-ichiro} Kishine}},\
  and\ \bibinfo {author} {\bibfnamefont {A.}~\bibnamefont {Ovchinnikov}},\
  }\href@noop {} {\bibfield  {journal} {\bibinfo  {journal} {Phys. Rev. B}\
  }\textbf {\bibinfo {volume} {78}},\ \bibinfo {pages} {064425} (\bibinfo
  {year} {2008})}\BibitemShut {NoStop}%
\bibitem [{\citenamefont {\mbox{Jun-ichiro} Kishine}\ and\ \citenamefont
  {Ovchinnikov}(2009)}]{Kishine09}%
  \BibitemOpen
  \bibfield  {author} {\bibinfo {author} {\bibnamefont {\mbox{Jun-ichiro}
  Kishine}}\ and\ \bibinfo {author} {\bibfnamefont {A.~S.}\ \bibnamefont
  {Ovchinnikov}},\ }\bibfield  {title} {\bibinfo {title} {Theory of spin
  resonance in a chiral helimagnet},\ }\href
  {https://doi.org/https://doi.org/10.1103/PhysRevB.79.220405} {\bibfield
  {journal} {\bibinfo  {journal} {Phys. Rev. B}\ }\textbf {\bibinfo {volume}
  {79}},\ \bibinfo {pages} {220405(R)} (\bibinfo {year} {2009})}\BibitemShut
  {NoStop}%
\bibitem [{\citenamefont {Kishine}\ and\ \citenamefont
  {Ovchinnikov}(2010)}]{Kishine10}%
  \BibitemOpen
  \bibfield  {author} {\bibinfo {author} {\bibfnamefont {J.}~\bibnamefont
  {Kishine}}\ and\ \bibinfo {author} {\bibfnamefont {A.}~\bibnamefont
  {Ovchinnikov}},\ }\bibfield  {title} {\bibinfo {title} {Sliding conductivity
  of a magnetic kink crystal in a chiral helimagnet},\ }\href@noop {}
  {\bibfield  {journal} {\bibinfo  {journal} {Phys. Rev. B}\ }\textbf {\bibinfo
  {volume} {82}},\ \bibinfo {pages} {064407} (\bibinfo {year}
  {2010})}\BibitemShut {NoStop}%
\bibitem [{\citenamefont {Kishine}\ \emph {et~al.}(2011)\citenamefont
  {Kishine}, \citenamefont {Proskurin},\ and\ \citenamefont
  {Ovchinnikov}}]{Kishine11}%
  \BibitemOpen
  \bibfield  {author} {\bibinfo {author} {\bibfnamefont {J.}~\bibnamefont
  {Kishine}}, \bibinfo {author} {\bibfnamefont {I.}~\bibnamefont {Proskurin}},\
  and\ \bibinfo {author} {\bibfnamefont {A.}~\bibnamefont {Ovchinnikov}},\
  }\bibfield  {title} {\bibinfo {title} {Tuning magnetotransport through a
  magnetic kink crystal in a chiral helimagnet},\ }\href@noop {} {\bibfield
  {journal} {\bibinfo  {journal} {Phys. Rev. Lett.}\ }\textbf {\bibinfo
  {volume} {107}},\ \bibinfo {pages} {017205} (\bibinfo {year}
  {2011})}\BibitemShut {NoStop}%
\bibitem [{\citenamefont {Shinozaki}\ \emph {et~al.}(2017)\citenamefont
  {Shinozaki}, \citenamefont {Hoshino}, \citenamefont {Masaki}, \citenamefont
  {Bogdanov}, \citenamefont {Leonov}, \citenamefont {Kishine},\ and\
  \citenamefont {Kato}}]{shinozaki2017fan}%
  \BibitemOpen
  \bibfield  {author} {\bibinfo {author} {\bibfnamefont {M.}~\bibnamefont
  {Shinozaki}}, \bibinfo {author} {\bibfnamefont {S.}~\bibnamefont {Hoshino}},
  \bibinfo {author} {\bibfnamefont {Y.}~\bibnamefont {Masaki}}, \bibinfo
  {author} {\bibfnamefont {A.}~\bibnamefont {Bogdanov}}, \bibinfo {author}
  {\bibfnamefont {A.}~\bibnamefont {Leonov}}, \bibinfo {author} {\bibfnamefont
  {J.-i.}\ \bibnamefont {Kishine}},\ and\ \bibinfo {author} {\bibfnamefont
  {Y.}~\bibnamefont {Kato}},\ }\bibfield  {title} {\bibinfo {title} {Fan-type
  spin structure in uni-axial chiral magnets},\ }\href@noop {} {\bibfield
  {journal} {\bibinfo  {journal} {arXiv preprint arXiv:1705.07778}\ } (\bibinfo
  {year} {2017})}\BibitemShut {NoStop}%
\end{thebibliography}%

\end{document}